\documentclass[11pt]{article}
\usepackage{graphics}
\usepackage{amsmath}
\usepackage{fancyhdr}
\usepackage{amssymb}
\usepackage{latexsym}
\usepackage{graphicx}
\usepackage{hyperref}
\usepackage{hyperref}
\hypersetup{backref, colorlinks=true}
\usepackage{epsfig}
\usepackage{setspace}
\usepackage{graphicx}
\usepackage{epstopdf}
\usepackage{subfigure}
\def\rz{\ifmmode{I\hskip -3pt R}
    \else{\hbox{$I\hskip -3pt R$}}\fi}
\def\nz{\ifmmode{I\hskip -3pt N}
    \else{\hbox{$I\hskip -3pt N$}}\fi}
\def\gz{\ifmmode{Z\hskip -4.8pt Z}
    \else{\hbox{$Z\hskip -4.8pt Z$}}\fi}
\def\cz{\ifmmode{C\hskip -4.8pt\vrule height5.8pt\hskip6.3pt}
\else{\hbox{$C\hskip -4.8pt\vrule height6.0pt\hskip6.3pt$}}\fi}
\def\qz{\ifmmode{Q\hskip -5.0pt\vrule height6.0pt depth0pt
   \hskip6pt}
    \else{\hbox{$Q\hskip -5.0pt\vrule height6.0pt depth0pt
   \hskip6pt$}}\fi}

\newcommand{\beq}{\begin{equation}}
\newcommand{\bseqs}{\begin{subequations}}
\newcommand{\eseqs}{\end{subequations}}
\newcommand{\balign}{\begin{align}}
\newcommand{\ealign}{\end{align}}
\newcommand{\eeq}{\end{equation}}
\newcommand{\beql}{\begin{equation} \label}
\newcommand{\beqs}{\begin{eqnarray}}
\newcommand{\eeqs}{\end{eqnarray}}
\newcommand{\beas}{\begin{eqnarray*}}
\newcommand{\eeas}{\end{eqnarray*}}
\newcommand{\ber}{\begin{array}}
\newcommand{\eer}{\end{array}}
\newcommand{\becs}{\begin{cases}}
\newcommand{\eecs}{\end{cases}}

\newcommand{\leftm}{\left[\begin{array}}
\newcommand{\rightm}{\end{array}\right]}

\newcommand{\bfa}{{\mathbf a}}
\newcommand{\bfb}{{\mathbf b}}

\newcommand{\bfd}{{\mathbf d}}
\newcommand{\bfe}{{\mathbf e}}
\newcommand{\bff}{{\mathbf f}}
\newcommand{\bfg}{{\mathbf g}}

\newcommand{\bfj}{{\mathbf j}}

\newcommand{\bfn}{{\mathbf n}}

\newcommand{\bfq}{{\mathbf q}}

\newcommand{\bfs}{{\mathbf s}}

\newcommand{\bfu}{{\mathbf u}}
\newcommand{\bfv}{{\mathbf v}}

\newcommand{\bfx}{{\mathbf x}}

\newcommand{\bfA}{{\mathbf A}}
\newcommand{\bfB}{{\mathbf B}}
\newcommand{\bfC}{{\mathbf C}}

\newcommand{\bfF}{{\mathbf F}}

\newcommand{\bfI}{{\mathbf I}}

\newcommand{\bfJ}{{\mathbf J}}

\newcommand{\bfL}{{\mathbf L}}
\newcommand{\bfM}{{\mathbf M}}
\newcommand{\bfN}{{\mathbf N}}

\newcommand{\bfQ}{{\mathbf Q}}
\newcommand{\bfR}{{\mathbf R}}
\newcommand{\bfS}{{\mathbf S}}

\newcommand{\ubf}{{\bf u}}

\newcommand{\xbf}{{\bf x}}

\newcommand{\Fbf}{{\bf F}}

\newcommand{\IIbf}{{\bf I\!I}}

\newcommand{\Qbf}{{\bf Q}}
\newcommand{\Rbf}{{\bf R}}

\newcommand{\dT}{{{\delta T}}}
\newcommand{\bbm}{\begin{bmatrix}}
\newcommand{\ebm}{\end{bmatrix}}

\newcommand{\bfbeta}{{{\boldsymbol \beta}}}

\newcommand{\bfkappa}{{{\boldsymbol \kappa}}}

\newcommand{\bfatld}{{\tilde{\bfa}}}

\newcommand{\sigmahat}{{\hat{\sigma}}}
\newcommand{\kappahat}{{\hat{\kappa}}}

\newcommand{\bfAtld}{{\tilde{\bfA}}}

\newcommand{\Wdot}{\dot{ W}}

\newcommand{\bfCtld}{\tilde{\bf C}}

\newcommand{\bfsigma}{{{\boldsymbol{\sigma}}}}

\newcommand{\dbfC}{{{\vartriangle}\bf C}}

\newcommand{\bbW}{{\mathbb{W}}}

\newcommand{\dbfA}{{{\vartriangle\!\!\bfA}}}

\newcommand{\bfepsilon}{{\boldsymbol{ \epsilon}}}

\newcommand{\intbar}{{\int\!\!\!\!\!- }} 
\newcommand{\inttbar}{{\int\!\!\!\!\!\!- }} 

\newcommand{\eps}{{\varepsilon}}

\newcommand{\half}{\frac{\;1}{\;2}}

\newcommand{\diverg} {{\rm div}}

\newcommand{\iie}{{{\rm i.e.}}}
\newcommand{\Tr}{{{\rm Tr}}}
\newcommand{\iif}{{{\rm if}}}

\newcommand{\oon}{{{\rm \;on\;}}}

\newcommand{\shat}{{\hat{s}}}

\begin{document}

\title{A continuum theory of thermoelectric bodies and effective  properties of thermoelectric composites}

 \vspace{0.5cm}

\author{\Large Liping Liu\\
{\em
Department of Mechanical Aerospace Engineering, Rutgers University, NJ 08854}\\
 {\em Department of Mathematics, Rutgers University, NJ 08854
}}
\date{ }
\maketitle

\begin{center}
 \vspace{0.2cm}
 Draft: March, 2012\\
 In press at {\em Internal Journal of Engineering Science}, 2012.
\end{center}

\tableofcontents
\newpage

\begin{center}

\section*{A continuum theory of thermoelectric bodies and effective \\ properties of thermoelectric composites}

\vspace{1cm}
\section*{Abstract}
\end{center}
We develop a continuum theory for thermoelectric bodies following the framework of continuum mechanics and conforming to general principles of thermodynamics. For steady states, the governing equations for local fields are intrinsically nonlinear. However, under conditions of small variations of electrochemical potential, temperature and their gradients, the governing equations can be reduced to a linear elliptic system and conveniently solved to determine local fields in thermoelectric bodies. The linear theory is further applied to  predict effective properties of thermoelectric composites. In particular, explicit formula of effective properties are obtained for simple microstructures of laminates and periodic E-inclusions, which imply useful design principles for engineering thermoelectric composites.
\\

\section{Introduction}
Thermoelectric (TE) materials directly convert heat into electric energy. They are often used for power generation and refrigeration (Rowe 1999; Mahan 2001). Being all solid state and without moving parts, thermoelectric devices have the unique advantage of portability, silence, scalability, durability and reliability in extreme environment (DiSalvo 1999). Also, they are readily customized and embedded into a large system to provide power or refrigeration. For example,  thermoelectric materials have been used to power independent wireless systems (Leonov {\em et al.}, 2008) and cool microelectronic chips (Boukai {\em et al.}, 2008; Rowe 2006). Additional applications of thermoelectric devices include temperature and  thermal energy sensors for precise temperature and thermal flux control (Riffat and Ma 2003),  hybrid photovoltaic-thermoelectric energy systems which have an improved solar power conversion efficiency (Kraemer {\em et al.}, 2007), etc. However, thermoelectric devices suffer from a critical disadvantage of low efficiency.

Recent renaissance of research in thermoelectrics has been focusing on improving thermoelectric efficiency, measured by a dimensionless figure of merit ($ZT_0$). Equipped with modern technologies of nanofrabrication, tremendous amount of efforts have been devoted to explore cost efficient methods for manufacturing TE materials with high figure of merit, see, e.g.,  Yang {\em et al.} (2000), Venkatasubramanian {\em et al.} (2001), Yamashita {\em et al.} (2005), Ohta {\em et al.} (2007), reviews of  Mahan (1998), Nolas {\em et al.} (2006), Snyder and Toberer (2008),  Lan et al. (2010) and references therein. An important method is to consider TE nanocomposites which has demonstrated potential of improving properties of TE materials including figure of merit $ZT_0$, power factor, structural integrity and adaptivity, etc. Therefore, the quest for desirable TE materials would benefit greatly from a rigorous homogenization theory which can predict the macroscopic properties of TE nanocomposites and furnish design strategies for desired effective properties. Such a homogenization theory above all demands a continuum theory for thermoelectric bodies. Moreover, on the device level the geometry and boundary condition on the constituent thermoelectric body can significantly influence local temperature and electrochemical potential on the body and hence the behavior of the device. The disparity between actual  geometry and boundary condition in a functional device and ideal geometry and environment under which TE properties are measured may  explain why the efficiency of TE devices is usually smaller than the efficiency of the  TE material (Snyder and Toberer 2008). It also calls into caution how to interpret experimental data when the geometries and boundary conditions of samples are significantly different from the ideal ones (Yamashita {\em et al.} 2005; Yamashita and Odahara 2007). To quantitatively account for the influence of  geometry and environment, we again need a self-consistent continuum theory capable of predicting all relevant local fields in thermoelectric bodies.

As a continuum theory, the presented model of thermoelectric bodies shall consist of three components: kinematics, conservation laws and constitutive law, and this model shall conform to general principles of thermodynamics. Kinematically we describe the local thermodynamic state in a thermoelectric body by  electrochemical potential and temperature $(\mu, T)$. The conservation laws can be formulated for energy flux $\bfj_u$, heat flux $\bfq$, electric flux (or currents) $\bfj_e$ or entropy flux $\bfj_s$, among which only two are independent. The constitutive law is usually postulated as a linear relation between fluxes and driving-forces which are typically gradient fields derived from state variables $(\mu, T)$. Such a constitutive law gives rise to a coefficient matrix $\bfL$ which may be measured by experiments or computed from microscopic theories (Ashcroft and Mermin 1976). The conformity to thermodynamic laws calls into question the validity of a postulated equation $\diverg \bfj_s=0$ for steady states, used by, e.g., Bergman and Levy (1991) since $\gamma=\diverg \bfj_s$ can be identified as the rate of entropy production per unit volume for steady states (Gurtin {\em et al.}, 2010). As a transport and irreversible process, a nontrivial steady-state of a thermoelectric body shall  generate entropy and dissipate energy, i.e., $\gamma>0$. Moreover, there exist certain reciprocal relations in the coefficient matrix $\bfL$  as implied by Onsager's principle which is sometimes referred to as the ``Fourth law of thermodynamics'' (Wendt 1974). However, reciprocal relations hold  only for ``proper, conjugate'' pairs of fluxes and driving-forces, which gives rise to many confusions in the literature  as for the ``most convenient'' pairs of fluxes and driving-forces. What make a bad situation worse lies in the temperature dependence of the coefficient matrix $\bfL$. At a large temperature scale, the coefficient matrix $\bfL$ should inevitably depend on temperature $T$, which would make the macroscopic theory nonlinear and untractable;  on the other hand we may assume the coefficient matrix $\bfL$ or any of its algebraic combinations with $T$ and $\mu$ remain constant on a homogeneous thermoelectric body if  variations of  $T$ and $\mu$ are small.

The above issues motivate us to scrutinize the underlying conditions or hypotheses for  material properties,  conservation laws,  constitutive law, and important derived material properties such as figure of merit and power factor.  We remark that  essential elements of  the proposed continuum theory are  well-known to some extent in earlier publications, e.g.,  Domenicali (1954), Callen (1948; 1960), Barnard (1972), de Groot and Mazur (1984), Haase (1990), Milton (2002). Here our focus is to develop the theory systematically following the framework of continuum mechanics. In particular, we clarify the conditions for applying self-consistently the proposed continuum theory: small variations of ($\mu, T$) and small driving-forces. The former condition is needed for assuming some algebraic combinations of the coefficient matrix and $(\mu, T)$ may be regarded as constant in a homogeneous TE body; the latter is needed for the postulated linear constitutive law and Onsager's reciprocal relations. These conditions are usually satisfied in practical applications of thermoelectric materials. If they were violated, a more complete model including temperature-dependence of material properties and nonlinear dependence of fluxes on driving-forces would be necessary.

An important feature of the proposed theory is that the boundary value problem for determining local fields is a linear system of partial differential equations whose existence, uniqueness and stability have been thoroughly investigated, see e.g. Evans (1998). This is convenient since a nonlinear set of partial differential equations are rarely solvable, meaning reliable predictions are difficult and rare. Moreover, the established homogenization methods can be used for this linear system and we obtain the unit cell problem for predicting effective properties of TE composites. For simple microstructures of laminates and periodic E-inclusions, the unit cell problem admits closed-form solutions, providing useful guide for designing TE composites.

The paper is organized as follows. We begin with experimental observations and clarify conditions for defining conventional thermoelectric properties in Section~\ref{sec:EO}.  In Section~\ref{sec:CL} we formulate the conservation laws for energy flux $\bfj_u$ and electric flux $\bfj_e$. In Section~\ref{sec:CM1} we identity pairs of ``proper, conjugate'' fluxes and driving-forces to apply Onsager's reciprocal relations, and explain the rationale of assuming  the coefficient matrix between chosen pairs of fluxes and driving-forces are constant on a homogeneous body. An alternative viewpoint of Onsager's reciprocal relations and positive-definiteness of the coefficient matrix is provided in Section~\ref{sec:CM2}. In Section~\ref{sec:BVP} we formulate  boundary value problems for various applied boundary  conditions which can be used to determine local fields in TE bodies, and solve the problems for an infinite plate and circular or spherical shell in Section~\ref{sec:FP1} and \ref{sec:FP2}, respectively. We proceed to define the effective properties of thermoelectric composites by the homogenization theory in Section~\ref{sec:EP0}. Explicit formula of effective TE properties are obtained for composites with microstructures of simple laminates in Section~\ref{sec:EPL} and of periodic E-inclusions in Section~\ref{sec:EPE}.  The applications of these closed-form formula to design desirable composites are explored in Section~\ref{sec:EPA}. We finally summarize in Section~\ref{sec:SD} and provide a derivation of Onsager's reciprocal relations for TE materials in Appendix A.

\renewcommand{\theequation}{\thesection.\arabic{equation}}
\setcounter{equation}{0}
\section{A continuum model for thermoelectric bodies}
\subsection{Experimental observations} \label{sec:EO}

We first describe properties of  thermoelectric materials observed in experiments. Consider a homogeneous   body $\Omega\subset \rz^n$, where $n=2\;{\rm or}\;3$ is the dimension of space. The thermodynamic state of the body is described by electrochemical potential $\mu$ and absolute temperature $T$. Further, we view as primitive concepts such as  internal energy density $u$, entropy density $\sigma$, electric flux $\bfj_e$,  entropy flux $\bfj_s$,  entropy supply $\eta$ (by external sources),  and the identities
\beqs \label{eq:bfjsbfq}
\bfq=T\bfj_s, \qquad r= T\eta,
\eeqs
where $\bfq$ is heat flux and $r$ is external heat supply inside the body (owing to radiation).  Note that these quantities are pointwisely defined inside the continuum body $\Omega$, permitted by the concept of {\em quasi-closed} and {\em local equilibrium} of subsystems in statistical mechanics (Landau and Lifshitz 1999, page 2). Suppose that there is no external heat supply inside the body, i.e., $r=0$; the only way of heat exchange with the environment is through
the boundary of $\Omega$ by conduction. For a time-independent boundary condition on $\partial \Omega$,  the body eventually evolves into a steady state for which the fluxes or the driving-forces, i.e.,  gradients  of
electrochemical potential and temperature, can be measured.
Experimentally, the electric conductivity tensor $\bfsigma\in \rz^{n\times n}$,  thermal conductivity tensor $\bfkappa\in \rz^{n\times n}$, Seebeck coefficient matrix $\bfs\in \rz^{n\times n}$  and Peltier coefficient matrix $\bfbeta\in \rz^{n\times n}$ are introduced by the following observations:
\begin{enumerate}
  \item At zero temperature gradient, i.e., $\bfg=-(\nabla T)^T=0$, the electric flux $\bfj_e$ is given by the {\em Ohm's Law}:
\beqs \label{eq:sigma} \bfj_\bfe=\bfsigma \bfe, \qquad\bfe=-(\nabla
\mu )^T\;\;\oon\;\Omega. \eeqs
Note that the driving-force field $\bfe$ includes the electric field
and chemical contribution from nonuniform electron concentration.

\item At zero electric flux, i.e., $\bfj_\bfe=0$ (open circuit), the heat flux $\bfq$ is given by the {\em Fourier's Law}:
\beqs \label{eq:kappa} \bfq=T\bfj_s=\bfkappa \bfg,\qquad \bfg=-(\nabla
T)^T\;\;\;\;\oon\;\Omega. \eeqs

\item At zero electric flux, i.e., $\bfj_e=0$ (open circuit), the driving-force field $\bfe=-(\nabla \mu)^T$ for an electric charge is given by
\beqs \label{eq:alpha} \bfe=-\bfs \bfg,\qquad \bfg=-(\nabla T)^T
\qquad \oon\;\Omega. \eeqs The above coupling between the field $\bfe$ and temperature gradient $\bfg$ is called the Seebeck effects.
  \item  At zero temperature gradient, i.e., $\bfg=-(\nabla T)^T=0$, the heat flux $\bfq$ is given by
  \beqs \label{eq:beta}
\bfq=T\bfj_s=\bfbeta \bfj_\bfe \qquad \oon\;\Omega.
  \eeqs
  The above coupling between the heat flux $\bfq$ and and electric flux $\bfj_e$ is called the Peltier effect.
\end{enumerate}

From \eqref{eq:sigma}-\eqref{eq:beta}, we infer  the following phenomenological  relation  between fluxes and
driving-forces: \beqs \label{eq:constituitive}
\bfj_1=\bfL_1\bff_1,\qquad
\bfj_1:=\begin{bmatrix}
\bfj_\bfe\\
\bfj_s\\
\end{bmatrix},\quad
\bff_1:=
\frac{1}{T} \begin{bmatrix}
\bfe\\
\bfg\\
\end{bmatrix}, \quad \bfL_1:=\begin{bmatrix}
T\bfsigma &T\bfsigma \bfs  \\
 \bfbeta \bfsigma&  (\bfkappa+ \bfbeta \bfsigma \bfs)\\
\end{bmatrix}.
\eeqs where  $\bfL_1\in \rz^{2n\times 2n}$ and  $\bfj_e,\; \bfj_s, \;\bfg,\; \bfe$ are understood as $n\times 1$ column vectors.
In general the coefficient matrix $\bfL_1$ shall depend on temperature $T$ and independent of electrochemical potential~$\mu$.  Further, by Thomson or Kelvin relations, we have $\bfL_1=\bfL_1^T$ or equivalently,
\beas
\bfsigma^T=\bfsigma, \quad \bfbeta=T\bfs^T, \quad \bfkappa^T=\bfkappa.
\eeas
 The above symmetry can be regarded as a consequence of Onsager's principle of microscopic reversibility (Onsager 1931 and Appendix). Further, from the Second law of thermodynamics, $\bfL_1$ is positive definite, and henceforth for any
  nonzero vectors $\bfv_1,\bfv_2\in \rz^n$,
 \beqs \label{eq:bfL1positive}
T\bfv_1\cdot \bfsigma \bfv_1+ 2 T \bfv_2 \cdot
 \bfsigma\bfs\bfv_1+\bfv_2\cdot(\bfkappa+T\bfs^T\bfsigma\bfs)\bfv_2>0.
 \eeqs

From a microscopic viewpoint the phenomenological linear relation \eqref{eq:constituitive} can be justified as the leading-order approximation when driving-forces on the carriers (i.e., electrons or holes) are weak (Ashcroft and Mermin 1970, chapter 13). In present context the driving-forces have two origins:  electrochemical potential gradient and temperature gradient. To compare them self-consistently, we shall assume (cf., equation (13.43) of Ashcroft and Mermin 1970)
\beqs \label{eq:weakforce}
\frac{k_B{\nabla T}}{F_0}\sim \frac{ {e \nabla \mu}}{F_0}\sim \eps<<1,
\eeqs
where $k_B=1.38\times 10^{-23} J/K$ is the Boltzmann constant, $e=1.60\times 10^{-19} C$ is the charge of an electron, and $F_0$ is the force such that the linear relation~\eqref{eq:constituitive} is no longer applicable (when carriers are subject to driving forces at the order of $F_0$). This force $F_0$ is a material's property and may be estimated from a pure electric measurement: $F_0\sim e E_0$, where $E_0$ is the critical electric field strength $|\bfe|$ such that the Ohm's Law $\bfj=\bfsigma \bfe$ is no longer sufficient to describe the flux-field relation.

\subsection{Conservation laws} \label{sec:CL}
Denote by $\rho$ the free charge density and $u$  the internal energy density. The conservation of electric charges implies
\beqs
\label{eq:econserved} \diverg \bfj_e =-\partial_t \rho \qquad \oon\;\Omega. \eeqs
Further, the internal energy density can be written as
\beqs \label{eq:u}
u=u_0+C_v T+\rho \mu,
\eeqs
where $u_0$ is a constant independent of $T, \mu$, and $C_v$ is the volumetric specific heat. Also, we identify energy flux $\bfj_u$ as
 \beqs \label{eq:jujsje}
\bfj_u=T\bfj_s+\mu \bfj_e,
\eeqs
and by conservation of energy, obtain
\beqs \label{eq:uconserved} \diverg \bfj_u=\bfj_s\cdot \nabla T+T\diverg \bfj_s+\bfj_e \cdot \nabla \mu+\mu \diverg \bfj_e=-\partial_t u \qquad
\oon\;\Omega. \eeqs

\subsection{A constitutive model  for thermoelectric materials} \label{sec:CM1}

There is a lot of confusion as to appropriate dependence of material properties  (i.e., the coefficient matrix $\bfL_1$ in \eqref{eq:constituitive}) on state variables ($\mu, T$) and how to justify the symmetry $\bfL_1^T=\bfL_1$ by Onsager's principle. From a simple dimension analysis, we find that the symmetry $\bfL_1=\bfL_1^T$ requires the dimensions of the two inner products between pairs of flux and driving-force, i.e., $\bfj_e\cdot \bfe$ and $\bfj_s\cdot \bfg$, should have the same dimension. Further, let $\Lambda\in \rz^{2n\times 2n}$ be an invertible matrix. A transformation of fluxes and driving-forces
\beqs \label{eq:transform}
(\bfj_1, \bff_1)\to (\bfj_2,\bff_2)=(\Lambda \bfj_1, \Lambda^{-T} \bff_1)
\eeqs
 implies
\beqs \label{eq:bfjbffp}
\bfj_2=\bfL_2\bff_2, \qquad \bfL_2=\Lambda\bfL_1\Lambda^{T}.
\eeqs
Therefore, there is no unique choice of fluxes and driving-forces for describing the transport process. Moreover, we can switch the roles played by the flux and driving-forces and, e.g.,  refer to $\bff_3:=[\bfj_e,\bfg/T]$ as the ``driving-forces'' and $\bfj_3:=[\bfe/T, \bfj_s]$ as the ``fluxes''. Then by \eqref{eq:constituitive} we have
\beqs \label{eq:transform2}
\bff_3:=
\bbm \bfj_e\\
\bfg/T\ebm, \quad
\bfj_3:=\bbm
\bfe/T\\ \bfj_s \ebm, \quad
\bfj_3=\bfL_3\bff_3,\quad
\qquad \bfL_3=
\bbm
\bfsigma^{-1}/T& -\bfs\\
\bfs^T& \bfkappa\\
\ebm,
\eeqs
where the off-diagonal blocks  in $\bfL_3$ are skew-symmetric instead of symmetric.  Therefore, the Onsager's reciprocal relations do not always imply the symmetry of the coefficient matrix; the symmetric property of the coefficient matrix depends on the choice of driving-forces and fluxes (Coleman and Truesdell 1960). To remedy this issue, we subsequently require driving-forces be  even functions of microscopic velocities of carriers whereas fluxes be odd functions of microscopic velocities of carriers,  and hence the coefficient matrix would be symmetric by the Onsager's reciprocal relations (Casimier 1945).

It is clear that the inner products of pairs of driving-forces and fluxes remain invariant for transformations in \eqref{eq:transform} and \eqref{eq:transform2}:
 \beqs \label{eq:dotinv}
 \bfj_1\cdot \bff_1=\bfj_2\cdot \bff_2=\bfj_3\cdot \bff_3.
 \eeqs
 Indeed, the fundamental thermodynamics for irreversible processes implies that the rate of entropy production per unit volume, denoted by $\gamma$,  cannot depend on the choice of fluxes and driving-forces used to describe the irreversible process, and  the conjugate pair of fluxes and driving-forces $(\bfj, \bff)$, by definition, satisfy
(see Appendix A)
\beqs  \label{eq:gamma0}
\gamma=\bfj\cdot \bff.
\eeqs
From Gurtin et al (2010, page 188) we have
\beqs  \label{eq:gamma}
\gamma=\partial_t \sigma+\diverg \bfj_s,
\eeqs
where $\sigma$ is the entropy density. By the First law of thermodynamics we have
\beqs \label{eq:FirstLaw}
\partial_t u(\bfx, t)=T \partial_t \sigma(\bfx,t)+\mu  \partial_t \rho(\bfx,t),
\eeqs
and by conservation laws \eqref{eq:econserved}-\eqref{eq:uconserved},
\beqs \label{eq:divjs}
T\diverg \bfj_s=-\partial_t u-\bfj_s\cdot \nabla T-\bfj_e\cdot \nabla \mu +\mu  \partial_t \rho.
\eeqs
Inserting  \eqref{eq:divjs} into \eqref{eq:gamma}  and by \eqref{eq:FirstLaw} we obtain
\beqs \label{eq:gammajf}
\gamma=(-\bfj_e\cdot \nabla \mu-\bfj_s\cdot \nabla T)/T.
\eeqs
Therefore, ($\bfj_e, \bfj_s$) and ($-\nabla \mu/T, -\nabla T/T$) are a conjugate pair of fluxes and driving-forces; the symmetry $\bfL_1=\bfL_1^T$ in \eqref{eq:constituitive} follows from Onsager's principle of microscopic reversibility and the fact that $\mu$ and $T$ are even functions of microscopic velocities of carriers. In Appendix A we outline a procedure of deriving Onsager's reciprocal relations for thermoelectric materials.

For steady states, it is more convenient to use electric flux and energy flux $(\bfj_e, \bfj_u)$  since they are divergence free by conservation laws \eqref{eq:econserved}-\eqref{eq:uconserved}. To find their conjugate driving-forces, by \eqref{eq:jujsje} we notice the  relation ($\bfI$ is the $n\times n$ identity matrix)
\beas
\bfj:=
\begin{bmatrix}
\bfj_e\\
\bfj_u\\
\end{bmatrix}=
\Lambda
\begin{bmatrix}
\bfj_e\\
\bfj_s\\
\end{bmatrix}
,\qquad \Lambda=\begin{bmatrix}
\bfI&0\\
\mu \bfI&T\bfI\\
\end{bmatrix},
\eeas
and hence, by \eqref{eq:transform} and \eqref{eq:bfjbffp},  the conjugate driving-forces and coefficient matrix between the fluxes $\bfj$ (i.e., $\bfj=\bfL \bff$) and driving-forces $\bff$ are given by
\beas
\bff:=\Lambda^{-T}\begin{bmatrix}
-({\nabla \mu})^T/{T}\\
-({\nabla T})^T/{T}\\
\end{bmatrix}=
\begin{bmatrix}
[\nabla (-{\mu}/{T})]^T\\
[\nabla (1/{T})]^T\\
\end{bmatrix},
\eeas
\beqs \label{eq:bfC}
\bfL(\mu, T):=
\begin{bmatrix}
T\bfsigma& \mu T \bfsigma+T^2  \bfsigma \bfs\\
 \mu T \bfsigma+T^2  \bfs^T \bfsigma& \mu^2 T \bfsigma+\mu T^2 (\bfs^T\bfsigma+\bfsigma \bfs)+T^2(\bfkappa+T\bfs^T\bfsigma \bfs) \\
\end{bmatrix},
\eeqs
respectively. 

 As is well-known, an arbitrary additive constant in electrochemical potential $\mu$ should have no physical consequence. This makes the appearance of $\mu$ in the coefficient matrix $\bfL(\mu, T)$ somewhat strange. Nevertheless, we recall that $\bfsigma, \;\bfs,\;\bfkappa$ are independent of $\mu$ (see discussions in Section~\ref{sec:CM2}). Upon straightforward calculations it can be shown that if $(-\mu/T, 1/T)$ satisfy the conservation laws \eqref{eq:econserved}-\eqref{eq:uconserved}:
 \beqs \label{eq:diveq1}
 \diverg[\bfL(\mu, T)\bff ]=
 \bbm
 -\partial_t \rho\\
 -\partial_t u
 \ebm,\qquad \bff=
 \bbm
 \nabla (-{\mu}/{T}) \\
\nabla (1/{T})
 \ebm ,
 \eeqs
 then  $(-(\mu+c)/T, 1/T)$ satisfy that for any constant $c\in \rz$,
\beqs \label{eq:diveq2}
 \diverg[\bfL(\mu+c, T)\bff' ]=
 \bbm
 -\partial_t \rho\\
 -\partial_t u-c\partial_t \rho
 \ebm,\qquad \bff^{'}=
 \bbm
 \nabla (-{\mu}/{T})-c\nabla (1/T)\\
\nabla (1/{T})
 \ebm.
 \eeqs
The above equation confirms that an additive constant in electrochemical potential $\mu$ indeed has no physical consequence.

From experiments or microscopic statistical mechanics, the coefficient matrix between fluxes and driving-forces in general depends on temperature. This dependence makes a boundary value problem based on conservation laws \eqref{eq:econserved}-\eqref{eq:uconserved} nonlinear and untractable.  This difficulty may be addressed by introducing an additional hypothesis that the deviation of the actual $(\mu, T)$ from their equilibrium values $(\mu_0, T_0)$ is small. For small bodies, this hypothesis is contained in the hypothesis of weak driving-forces which is required by  the very linear relation between fluxes and driving-forces. Moreover, the Onsager's reciprocal relations require the system does not deviate too much from the equilibrium state which overlaps with  the hypothesis of small variations of  $(\mu, T)$. Therefore, in typical situations it is reasonable to assume that the coefficient matrix is  constant, independent of positions inside the homogeneous body $\Omega$, and given by
\beqs \label{eq:bfC0}
\bfL(\mu(\bfx), T(\bfx))=\bfL(\mu_0, T_0) \qquad \forall\,\bfx\in \Omega,
\eeqs
where $(\mu_0, T_0)$ is the equilibrium electrochemical potential and temperature of the body $\Omega$ when the body is closed and isolated. In practice we can choose $T_0$ as the average temperature of the body and $\mu_0=0$.

 We remark that equally sound but not equivalent is to assume the coefficient matrix between another pair of conjugate fluxes and driving-forces, e.g., the matrix $\bfL_1$ in \eqref{eq:constituitive}, is uniform inside the homogeneous body. Compared with \eqref{eq:bfC0}, this new  assumption would imply different solutions  to a specific problem, but the difference is presumably small and nonessential because of the same premises: weak driving-forces and small variations of $(\mu, T)$. However, as one will see below, equation~\eqref{eq:bfC0} implies an easy-to-solve linear system of equations  whereas $\bfL_1$ being constant implies a nonlinear system (Yang {\em et al.}, 2012).

The temperature dependence of electric and thermal conductivities and Seebeck coefficient matrix implied by~\eqref{eq:bfC0} is
\beqs \label{eq:Tdependence}
\bfsigma \propto \frac{1}{T},  \quad \bfkappa \propto \frac{1}{T^2}, \quad \bfs \propto \frac{1}{T}.
\eeqs
Ideally speaking, there could be thermoelectric materials with the above temperature dependence precisely over some temperature range.  Even for such materials, the $\mu$-dependence of the coefficient matrix in \eqref{eq:bfC} would still make \eqref{eq:diveq1} nonlinear for steady states. From this viewpoint, a rigorous continuum theory for thermoelectric bodies are intrinsically nonlinear.

Finally we notice that the first and second of \eqref{eq:Tdependence} may be justified by  a microscopic theory for metals at a temperature much higher than the Debye temperature, see e.g. equations (26.48) and (13.58) of  Ashcroft and Mermin (1976), whereas the justification of the last of \eqref{eq:Tdependence} appears to be elusive, see comments in page 258 of Ashcroft and Mermin (1976). We stress here that the reasoning for~\eqref{eq:bfC0} lies in the conditions of weak driving-forces and small variations of $\mu, T$ instead of an underlying microscopic theory, and that  equation~\eqref{eq:bfC0} can be applied to materials whose properties violate \eqref{eq:Tdependence}, e.g., semiconductors, as long as the assumed conditions are satisfied.

\subsection{An alternative viewpoint of the constitutive model} \label{sec:CM2}
 The Second law implies that the rate of entropy production per unit volume $\gamma=\bff\cdot\bfL\bff>0$ for an irreversible process, which implies the positive-definiteness of the coefficient matrix $\bfL$ in \eqref{eq:bfC}. Further, if the driving-forces $\bff=0$, i.e., the electrochemical potential and temperature are uniform on the body, the body is in the equilibrium state and hence $\gamma=0$, attaining its minimum value. These observations motivate us to {\em postulate} the constitutive law between  fluxes and driving-forces in irreversible processes by specifying the functional dependence of the entropy product rate  per unit volume $\gamma$  on state variables and driving-forces. Below we   derive the phenomenological linearity between fluxes and driving-forces,  Onsager's reciprocal relations and  positive definiteness of the coefficient matrix as  consequences of this postulation and  weak driving-forces.

 In analogy with a reversible process, e.g., the elastic theory of deformable bodies, and with the entropy product rate  per unit volume $\gamma $  playing the role of internal energy density,   we {\em postulate} that $\gamma $ is a smooth function of thermodynamic variables and  driving-forces. The driving-forces, by definition,  characterize the degree of deviation of a  local  subsystem from its equilibrium state. We consider  a generic  situation where the thermodynamic state  is described by $m$ variables: $\bfu=(u_1,\cdots, u_m)$ whose values would be uniform on the body in the equilibrium state and are even functions of microscopic velocity of carriers. Then the gradients $\bfF=\nabla \bfu \in \rz^{m\times n}$ would be the appropriate quantities characterizing local deviation from local equilibrium state, and the above postulation can be expressed as
\beas
\gamma =\gamma (\bfu, \bfF).
\eeas
Under the condition of weak driving-forces, i.e., $|\bfF|<<1$, we have the Taylor expansion
 \beqs \label{eq:Wexpand0}
\gamma (\bfu,\bfF)=\gamma _0(\bfu)+\bfB(\bfu)\cdot \bfF+ \bfF\cdot \bfC(\bfu)
\bfF+\cdots, \eeqs
where \beqs \label{eq:Cdef}
\gamma_0(\bfu)=\gamma(\bfu,
0),\quad \bfB(\bfu)=\frac{\partial \gamma(\bfu, \bfF)}{\partial
\bfF}\bigg|_{\bfF=0}, \quad  \bfC(\bfu)=\half \frac{\partial^2 \gamma(\bfu,
\bfF)}{\partial \bfF \partial \bfF}\bigg|_{\bfF=0}. \eeqs
Since $\gamma (\bfu, \bfF) >0$ if $\bfF\neq 0$ (irreversible process) and $\gamma (\bfu,0)=0$ if $\bfF=0$ (equilibrium state), we conclude that
 \beqs \label{eq:Wzero}
 \gamma_0(\bfu)=0, \quad \bfB(\bfu)=0, \quad
\bfC(\bfu)\mbox{ is symmetric and positive definite}. \eeqs

From the definition, the proper, conjugate fluxes of the driving-forces $\bfF$ are the quantities $\bfJ$ such that
\beqs \label{eq:bfJbfF}
\bfJ\cdot \bfF= \gamma(\bfu, \bfF)=\bfF\cdot \bfC(\bfu) \bfF,
\eeqs
where we have neglected higher order terms in \eqref{eq:Wexpand0}. 
A natural solution to the above equation is given by
\beqs \label{eq:conjugate}
(\bfJ)_{pi}=(\bfC)_{piqj}(\bfF)_{qj},\quad\iie,\quad\;\bfJ=\bfC(\bfu) \bfF.
\eeqs
Therefore, we conclude that the postulation that the entropy production rate per unit volume $\gamma$ is a smooth function of the driving-forces, to the leading order, implies the linear and reciprocal relations between the conjugate pairs of fluxes and driving-forces.

When specified to thermoelectric bodies, if we choose  $\bfu=(\mu, T)$ as the state variables  $\bfF=\nabla \bfu$ the driving-forces, we anticipate $\gamma(\bfu,\bfF)$ and hence the associated fourth order tensor $\bfC(\bfu)$ is independent of $\mu$, which  by \eqref{eq:gammajf} implies the coefficient matrix $\bfL_1$ in \eqref{eq:constituitive} must be independent of $\mu$. If we choose $\bfu=(u_1, u_2)=(-\mu/T, 1/T)$ as the state variables, $\nabla \bfu$ as the driving-forces, then by \eqref{eq:bfjbffp}, \eqref{eq:gammajf}, and \eqref{eq:bfC}  we identify
\beas
\bfF=\bbm
\nabla u_1\\
\nabla u_2\\
\ebm, \qquad
\bfJ=\bbm
\bfj_e^T\\
\vspace{-0.4cm}
\\
\bfj_u^T\\
\ebm,\qquad
\bfJ=\bfC(\bfu)\bfF,
\eeas
where the associated fourth-order tensor $\bfC(\bfu)\in \rz^{2\times n\times 2\times n}$ is given by
\beqs \label{eq:bfccomp0}
(\bfC)_{1i1j}=T(\bfsigma)_{ij}, \quad (\bfC)_{1i2j}=(\bfC)_{2j1i}=\mu T(\bfsigma)_{ij}+T^2 (\bfsigma\bfs)_{ij}, \nonumber\\
(\bfC)_{2i2j}= [\mu^2 T \bfsigma+\mu T^2 (\bfs^T\bfsigma+\bfsigma \bfs)+T^2(\bfkappa+T\bfs^T\bfsigma \bfs)]_{ij}.
\eeqs
Further, for steady states, the conservation law \eqref{eq:diveq1} can be conveniently rewritten as
\beqs \label{eq:diveq3}
\diverg[\bfC(\bfu) \nabla \bfu]=0\qquad \oon\;\Omega,
\eeqs
where the divergence is taken over row vectors.

We remark that the above viewpoint also facilitates deriving the implications of material symmetry on material properties. For example, for isotropic materials we have $\gamma(\bfu, \bfF)=\gamma(\bfu, \bfF\bfR)$ for any rigid rotation $\bfR$, which, together with \eqref{eq:Wexpand0}, implies $\bfsigma$, $\bfkappa$ and $\bfs$ are isotropic matrix. A detailed discussion in a classic context can be found in Gurtin {\em et al.} (2010, \S 50).

\subsection{Boundary value problems for  thermoelectric bodies in steady states} \label{sec:BVP}
We now consider a heterogeneous thermoelectric body $\Omega$ whose material properties may in general depends on position.  If the boundary conditions applied to the body are time independent, the body would eventually evolve into a steady state. Suppose that the driving-forces to the carriers and variations of $(\mu, T)$ are small in  the overall body, and  we assume~\eqref{eq:bfC0} such that the thermoelectric tensor of this body is given by $\bfC(\bfu_0, \bfx)$, where $\bfu_0=(-\mu_0/T_0, 1/T_0)$ and $(\mu_0, T_0)$ are electrochemical potential and temperature of the body in the equilibrium state. We remark that  $(\mu_0, T_0)$ are uniform on $\Omega$ even if the body is heterogeneous. Therefore, upon choosing $\mu_0=0$ and setting $\bfu=(u_1, u_2)=(-\mu/T, 1/T)$, by \eqref{eq:diveq1} and \eqref{eq:diveq3} we infer that in steady states, the electrochemical potential and temperature fields satisfy
\beqs \label{eq:diveq4}
\becs
\diverg[\bfC_0( \bfx) \nabla \bfu]=0&\oon\;\Omega,\\
\mbox{time-independent boundary conditions} &\oon\;\partial \Omega,
\eecs
\eeqs
where
\beqs \label{eq:bfccomp00}
&&[\bfC_0(\bfx)]_{1i1j}=T_0[\bfsigma(\bfx)]_{ij}, \qquad [\bfC_0(\bfx)]_{1i2j}=[\bfC_0(\bfx)]_{2j1i}=T_0^2 [\bfsigma(\bfx)\bfs(\bfx)]_{ij},  \nonumber  \\
&&[\bfC_0(\bfx)]_{2i2j}= T_0^2[\bfkappa(\bfx)+T_0\bfs^T\bfsigma(\bfx) \bfs(\bfx)]_{ij},
\eeqs
and  the position-dependent $\sigma(\bfx), \bfs(\bfx)$, and $\bfkappa(\bfx)$ reflect that the body may be heterogeneous.
Further, let $\Gamma\subset \partial \Omega$ be a sub-boundary of the body $\Omega$ and $g, h$ be two given functions on $\Gamma$.
Typical boundary conditions on $\Gamma$ in applications  include the following.
\begin{enumerate}
  \item Dirichlet boundary condition: 
  \beqs \label{eq:bc11}
  T=g,\quad \mu=h \qquad \oon\;\Gamma,
  \eeqs
  where $g, h\;$ represents the temperature and electrochemical potential on $\Gamma$, respectively.
    \item Neumann boundary condition:
    \beqs \label{eq:bc22}
    \bfq\cdot \bfn=g, \quad \bfj_e\cdot\bfn=h \qquad \oon\;\Gamma,
    \eeqs
    where $\bfn$ is the outward normal on $\Gamma$, and $g, h\;$ represents the outward heat flux and electric flux on $\Gamma$, respectively.
  \item Mixed boundary condition: 
      \beqs \label{eq:bc33}
    L_1(\bfq\cdot \bfn, T)=g, \quad L_2(\bfj_e\cdot\bfn, \mu)=h \qquad \oon\;\Gamma,
    \eeqs
    where $L_1, L_2$ represent  linear functions of two variables.
\end{enumerate}
Other types of boundary conditions are also allowed from the mathematical viewpoint but physically less common.

A few remarks are in order here regarding transient states and sharp interfaces. Mahan (2000) suggested that for transient states the right hand sides of the conservation laws~\eqref{eq:econserved} and \eqref{eq:uconserved} should be given by
\beas
\partial_t\rho=0, \qquad \partial_t u=C_v \frac{\partial T}{\partial t},
\eeas
where $C_v$ is the heat capacity per unit volume. The above equations, though commonly assumed when there is no coupling between heat flux and electric flux, are questionable for heterogeneous thermoelectric bodies since there could be accumulations of free charges in transient states and not all energy flow into a region would be converted into heat for a thermoelectric body. To close the system of equations and in analogy with electrostatics, we introduce the field of local electric displacement $\bfd$ and postulate an additional constitutive law:
\beas
\diverg \bfd= \rho, \qquad \bfd=- \bfepsilon(\bfx)\nabla \mu,
\eeas
where $\bfepsilon$ may be interpreted as the ``permittivity'' of the thermoelectric body. Then by \eqref{eq:u} and conservation laws \eqref{eq:econserved}, \eqref{eq:uconserved} we obtain the following equations for transient states:
\beas
\diverg [\bfC_0(\bfx) \nabla \bfu(\bfx, t)]=\bbm
 \partial_t[\diverg \bfepsilon(\bfx) \nabla \mu]\\
-C_v\partial_t T+ \partial_t(\mu\diverg \bfepsilon(\bfx) \nabla \mu)
\ebm,
\;
\eeas
which appears to be a nonlinear system and difficult to solve.  Moreover, it was suggested that sharp interfaces might play an important role  in overall thermoelectric behaviors (Yamashita and Odahara 2007; Ohta {\em et al} 2007), which may be modeled by postulating a similar constitutive relation between fluxes and interfacial driving forces. Below we will not consider such interfacial effects for thermoelectric composites.


\renewcommand{\theequation}{\thesection.\arabic{equation}}
\setcounter{equation}{0}
\section{Power factor and figure of merit of thermoelectric structures}
In this section we solve boundary value problems  for simple structures of an infinite plate and a spherical shell and extract some combinations of basic material properties, namely, power factor $P_f$ and figure of merit $ZT_0$, for evaluating the capacity (per unit volume)  and  efficiency of power generation and refrigeration.

For more complex structures or boundary conditions, we anticipate  that the capacity and efficiency of power generation and refrigeration depend not only on material properties but also the overall geometry and boundary condition, which can be predicted by solving the proposed boundary value problem described in Section~\ref{sec:BVP}. An  interesting problem  to investigate is whether there exist geometries and boundary conditions such that the overall capacity (per unit volume) and efficiency of power generation and refrigeration exceed those of an infinite plate described below, and if not, why.

\subsection{A thermoelectric plate} \label{sec:FP1}

\begin{figure}[t]
\centering
\includegraphics[width=5.5in]{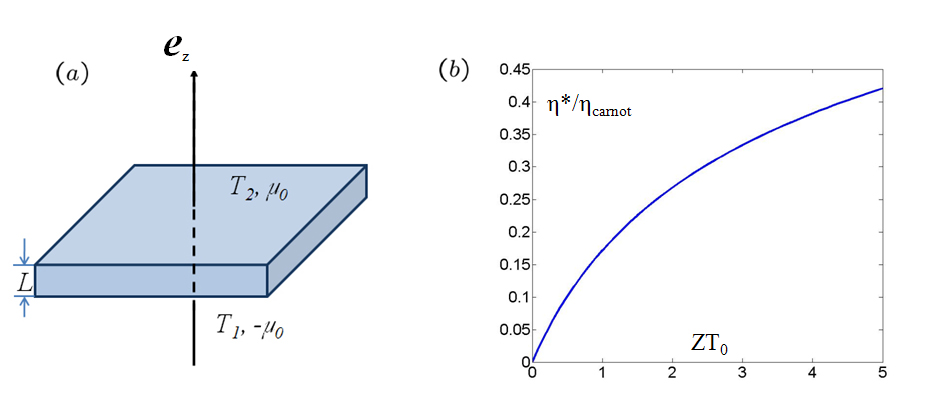}
\newline
\caption{ (a) An infinite thermoelectric plate of thickness $L$, electrochemical potential and temperature  $(T_1,-\mu_0)$ on the bottom face and $(T_2,\mu_0)$  on the top face; (b) normalized maximum efficient $\eta^\ast/\eta_{\rm carnot}$ versus  figure of merit $ZT_0$.}
\label{fig:1}
\end{figure}
 Consider a homogeneous and generally anisotropic infinite thermoelectric plate of thickness $L$ as shown in Fig.~\ref{fig:1} (a).  The temperature and electrochemical potential on top and bottom faces are maintained as
\beqs \label{eq:BC0}
T(z=0)=T_1, \quad T(z=L)=T_2, \quad \mu(z=0)=-\mu_0, \quad \mu(z=L)=\mu_0.
\eeqs
A priori we shall assume the temperature difference $\dT=T_2-T_1$ is small compared with the equilibrium temperature $T_0$, and hence
\beqs \label{eq:Tdiff}
\frac{1}{T_2}+\frac{1}{T_1}\approx \frac{2}{T_0}, \qquad \frac{1}{T_2}-\frac{1}{T_1}\approx -\frac{\dT}{T_0^2}.
\eeqs
Also, we assume the variation of electrochemical potential is small  and enforce the assumption~\eqref{eq:bfC0}.
Let $\bfe_z\in \rz^n$ be the unit normal to the plate as shown in Fig.~\ref{fig:1} (a) and
denote by
\beqs \label{eq:sigmabfe1}
\sigma_z= (\bfC)_{1i1j} (\bfe_z)_i(\bfe_z)_j/T_0, \quad \alpha_z=(\bfC)_{1i2j} (\bfe_z)_i(\bfe_z)_j/T_0^2,\quad
\kappa'_z=(\bfC)_{2i2j} (\bfe_z)_i(\bfe_z)_j/T_0^2,
\eeqs
which, by \eqref{eq:bfccomp00}, can be identified as
\beqs \label{eq:sigmabfe2}
\sigma_z= \bfe_z\cdot \bfsigma \bfe_z, \quad \alpha_z=\bfe_z\cdot \bfsigma \bfs \bfe_z, \quad \kappa'_z=\bfe_z\cdot (\bfkappa+T_0 \bfs^T\bfsigma \bfs)\bfe_z.
\eeqs
 Let
\beqs \label{eq:ZTPF}
E_f=\frac{\alpha_z^2}{\sigma_z\kappa'_z} T_0,\qquad P_f=\frac{\alpha_z^2}{\sigma_z}
\eeqs
be the {\em efficiency factor} and the {\em power factor} of the thermoelectric material. From \eqref{eq:bfL1positive}, choosing $\bfv_1=a \bfe_z$ and $\bfv_2=b \bfe_z$ for any $a, b\in \rz$, we see that
\beas
\sigma_z>0, \quad \kappa'_z>0, \quad  E_f<1.
\eeas
For isotropic materials, the above parameters $E_f$ and $P_f$ are independent of directional vector $\bfe_z$ and given by
 \beqs \label{eq:EfPf}
E_f=\frac{ T_0 s^2\sigma}{\kappa+ T_0 s^2\sigma}=\frac{ZT_0}{1+ZT_0},\qquad \qquad P_f=s^2 \sigma, 
 \eeqs
 where $ZT_0=T_0s^2\sigma/\kappa$ is the  {\em figure of merit} of thermoelectric materials. By the first of \eqref{eq:EfPf} we have
 \beqs \label{eq:ZT0}
 ZT_0=\frac{E_f}{1-E_f},
 \eeqs
 which remains valid for general anisotropic materials.  Below we explore the physical meanings of $ZT_0$ and $P_f$.

By  the  invariance of the problem for any translations within the plate, we infer that the solution to \eqref{eq:diveq4} and \eqref{eq:BC0} satisfies
\beas
&&\bfj_e=-T_0 [\frac{\mu_0}{T_2L}-\frac{-\mu_0}{T_1L}]\bfsigma \bfe_z+ T_0^2 [\frac{1}{T_2L}-\frac{1}{T_1L}] \bfsigma\bfs \bfe_z,\\
&&\bfj_u=-T_0^2 [\frac{\mu_0}{T_2L}-\frac{-\mu_0}{T_1L}]\bfs^T\bfsigma \bfe_z+ T_0^2 [\frac{1}{T_2L}-\frac{1}{T_1L}](\bfkappa+T_0 \bfs^T\bfsigma \bfs) \bfe_z,
\eeas
and  hence  the normal electric and energy fluxes along $\bfe_z$-direction is given by
\beqs \label{eq:jeju}
&&j_e=\bfj_e\cdot \bfe_z=-\frac{T_0\mu_0 \sigma_z }{L}(\frac{1}{T_2}+\frac{1}{T_1})+\frac{T_0^2\alpha_z}{L}(\frac{1}{T_2}-\frac{1}{T_1}) \approx -\frac{2\mu_0 \sigma_z }{L}-\frac{\dT \alpha_z}{L},\\
&&j_u=\bfj_u\cdot \bfe_z=-\frac{T_0^2\mu_0 \alpha_z }{L}(\frac{1}{T_2}+\frac{1}{T_1})+\frac{T_0^2\kappa'_z}{L}(\frac{1}{T_2}-\frac{1}{T_1})\approx
-\frac{2T_0\mu_0 \alpha_z}{L}-\frac{\dT\kappa'_z}{L}. \nonumber
\eeqs
Therefore,  the rate of work per unit volume done by the plate,   the heat flux into the plate from the bottom  and the heat flux out of the plate from the top is given by
\beqs \label{eq:q1q2}
\Wdot=2 j_e \mu_0/L,\quad q_1=j_u+j_e\mu_0, \qquad q_2=j_u-j_e\mu_0,
\eeqs
respectively. Let
\beqs \label{eq:rho}
 \rho=\frac{\dT \alpha_z}{2\mu_0\sigma_z}
 \eeqs
 be a dimensionless number for future convenience.
By \eqref{eq:jeju}, \eqref{eq:ZTPF} and \eqref{eq:rho} we find that
\beqs \label{eq:jejurho}
j_e \mu_0\approx-\frac{2\mu_0^2\sigma_z}{L}(1+\rho), \qquad j_u\approx -\frac{4\mu_0^2\sigma_z}{L}\frac{T_0}{\dT}(\rho+\frac{\rho^2}{E_f}).
\eeqs

\subsubsection*{Power generation}
Without loss of generality we assume $\mu_0>0$.
If $q_1\ge q_2$, i.e., $j_e\ge 0\;\Leftrightarrow \rho\le -1$, the thermoelectric plate converts heat into electric energy.
The  electric power generated per unit volume  is given by
 \beas
\Wdot:=2\mu_0 j_e/L=-\frac{2}{L^2}(2\mu_0^2 \sigma_z+ \mu_0 \dT  \alpha_z).
\eeas
Therefore, for fixed applied temperature difference $\dT$ the maximum power generated per unit volume is given by
\beas
\Wdot_{\max}=\frac{\alpha_z^2 \dT^2}{4\sigma_z L^2}=\frac{P_f}{4}(\frac{ \dT}{ L})^2\qquad \mbox{at }\quad \rho=-2.
\eeas
We remark that  power factor $P_f$ is more important than efficiency (figure of merit $ZT_0$) for a power generation system where heat sources are abundant and cost is negligible (Rowe 1999; Narducci 2011).

Further, the efficiency of power generation  is given by
\beas
\eta_g=\frac{q_1-q_2}{\max\{|q_1|, |q_2|\}}=\frac{2\mu_0j_e}{|j_u|+\mu_0j_e}\approx
\frac{2}{ \frac{2T_0}{|\dT|}|(\rho+\frac{\rho^2}{E_f})/(1+\rho)|+1}.
\eeas
Upon maximizing the efficiency against $\rho\le -1$, we find
\beqs \label{eq:etagast1}
\eta_g^\ast=
\frac{\dT}{T_0}\frac{\sqrt{1+ZT_0}-1}{\sqrt{1+ZT_0}+1+\dT (\sqrt{1+ZT_0}-1)/(2T_0) },
\eeqs
which is achieved at  $\rho=-1-1/\sqrt{1+ZT_0}$.
We remark that since $\dT/T_0<<1$, the the above formula is well approximated by
\beqs \label{eq:etagast2}
\eta_g^\ast \approx
\frac{\dT}{T_0}\frac{\sqrt{1+ZT_0}-1}{\sqrt{1+ZT_0}+1 },
\eeqs
 which agrees with the optimal efficiency predicted by  Ioffe (1957, page 40).


\subsubsection*{Refrigeration}
Without loss of generality we assume $\dT>0$.
If $q_1\ge 0$, i.e.,
\beqs \label{eq:q1ge0}
q_1=-\frac{2}{L} [\sigma_z \mu_0^2+(T_0+\dT/2 )\alpha_z\mu_0 +  \frac{ \dT }{2  }\kappa'_z]\ge 0,
\eeqs
the thermoelectric plate extracts heat from the low temperature side and ejects heat into the high temperature side. Since $\sigma_z>0$, the above inequality is possible if and only if
\beqs \label{eq:dtineq1}
(T_0+\dT/2)^2 \alpha_z^2-2\sigma_z\kappa'_z \dT\ge 0\quad
\Rightarrow \;\;(1+\frac{\dT}{2T_0})^2-\frac{2}{E_f}\frac{\dT}{T_0}\ge 0.
\eeqs
Therefore,
\beqs \label{eq:dtineq2}
0<\dT\le \dT_{\max}:= 2T_0\frac{\sqrt{1+ZT_0}-1}{\sqrt{1+ZT_0}+1 },
\eeqs
which is consistent with the prediction of Goldsmid (2010, page 11) when $ZT_0$ is small and such that $\dT/T_0<<1$.
We also remark that  other possibilities implied by \eqref{eq:dtineq1} are omitted for contradicting our assumption $\dT/T_0<<1$. For the same reason, $\dT_{\max}$, the maximum temperature difference possibly maintained by a thermoelectric cooler,  cannot serve as a good measurement of  $ZT_0$ if $\dT_{\max}/T_0$ is not a small number. In this case, a more detailed analysis accounting for the temperature dependence of material properties is necessary to predict $\dT_{\max}$.

Further, by \eqref{eq:jejurho} the cooling efficiency is given by
\beas
\eta_c =\frac{q_1}{-2j_e\mu_0}= \frac{-\frac{2T_0}{\dT}(\frac{\rho^2}{E_f}+\rho)-(1+\rho)}{2(1+\rho)},
\eeas
where $\rho> -1$ since $j_e\mu_0<0$, meaning work is done to the thermoelectric plate. Upon maximizing $\eta_c$ against $\rho>-1$ we find
\beqs \label{eq:etacast}
\eta_c^\ast=\frac{T_0}{\dT} \bigg[\frac{\sqrt{1+ZT_0}-1}{\sqrt{1+ZT_0}+1 }-\frac{\dT}{2T_0} \bigg]\qquad {\rm at}\; \; \rho=-1+1/\sqrt{1+ZT_0},
\eeqs
which agrees with the prediction of Goldsmid (2010, page 10) when $\dT/T_0<<1$.

Recall that the ideal efficiency for a reversible system working between two heat reservoirs at temperature $T_1$ and $T_2$ is given by the Carnot efficiency:
\beas
\eta_{\rm carnot}\approx \frac{\dT}{T_0} \quad\mbox{ for power generation} \qquad \approx \frac{T_0}{\dT} \quad \mbox{for refrigeration},
\eeas
where the temperature difference $\dT=|T_2-T_1|<< T_0=\half(T_1+T_2)$.
Therefore, a common factor between the optimal efficiency and Carnot efficiency can be extracted out from \eqref{eq:etagast2} and \eqref{eq:etacast}:
\beas
\frac{\eta^\ast}{\eta_{\rm carnot}}=\frac{\sqrt{1+ZT_0}-1}{\sqrt{1+ZT_0}+1 },
\eeas
whose functional dependence on the figure of merit $ZT_0$ are shown in Fig.~\ref{fig:1} (b).

\begin{figure}[t]
\centering
\includegraphics[width=5.5in]{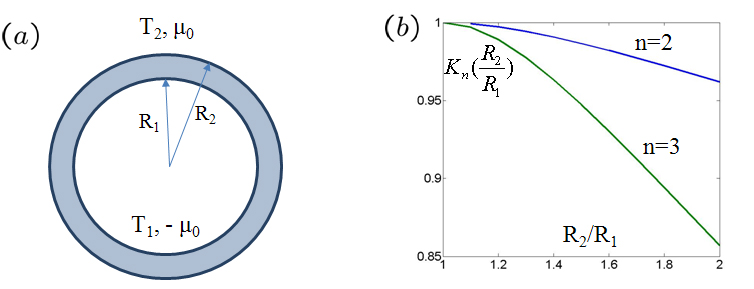}
\newline
\caption{ (a) A thermoelectric shell with prescribed temperatures and electrochemical potentials on the two faces; (b) the geometric factor $K_n(R_2/R_1)$ versus $R_2/R_1$ in \eqref{eq:Kn}.}
\label{fig:shell}
\end{figure}
\subsection{A thermoelectric shell}  \label{sec:FP2}
In applications it can be advantageous to use geometries other than plates/films for structural adaptivity, ease of thermal insulation or conduction, etc. In this section we explore the geometric effect by considering  spherical shells in two dimensions and three dimensions as shown in Fig.~\ref{fig:shell}~(a). For simplicity, we assume the thermoelectric material is isotropic with electric conductivity $\sigma$, Seekback coefficient $s$ and thermal conductivity $\kappa$, and the boundary conditions are given by
\beqs \label{eq:BC1}
T(r=R_1)=T_1, \quad T(r=R_2)=T_2, \quad \mu(r=R_1)=-\mu_0, \quad \mu(r=R_2)=\mu_0.
\eeqs
Again we assume the temperature difference $\dT=T_2-T_1$ is small compared with the equilibrium temperature $T_0$, and hence \eqref{eq:Tdiff} holds.
Below we derive the power factor and efficiency of this thermoelectric shell.

Since the material is isotropic, equation \eqref{eq:diveq4} can be written as
\beas
\becs
\Delta u_1+T_0s\Delta u_2=0  &\oon\;\Omega,\\
s\sigma \Delta u_1+(\kappa+T_0\sigma s^2) \Delta u_2=0&\oon\;\Omega.
\eecs
\eeas
From the symmetry we seek a solution to the above equations which can be written as $\bfu=(u_1(r), u_2(r))$.  Direct calculations show that
\beas
&&\frac{1}{r^{n-1}}\frac{d}{dr} r^{n-1}\frac{d}{dr}(u_1+T_0su_2)=0, \qquad \forall\,r\in (R_1, R_2), \\
&&\frac{1}{r^{n-1}}\frac{d}{dr} r^{n-1}\frac{d}{dr}(s\sigma u_1+(\kappa+T_0s^2\sigma )u_2)=0, \qquad \forall\,r\in (R_1, R_2).
\eeas
The general solution to the above equations is given as follows:
\beas
\iif\;n=2,\quad
\becs
u_1= -A_1\log(r)+A_0,\\
u_2=-B_1\log(r) +B_0;\\
\eecs
\qquad
\iif\;n=3,\quad
\becs
u_1= \frac{A_1}{ r}+A_0,\\
u_2=\frac{B_1}{r} +B_0,\\
\eecs
\eeas
where $A_1, A_0, B_1$ and $B_0$ are constants.
By boundary conditions~\eqref{eq:BC1}  and \eqref{eq:Tdiff} we find
\beqs \label{eq:AB}
\iif\;n=2, \quad
\becs
\vspace{0.2cm}
A_1 \approx \frac{2\mu_0}{T_0\log(R_2/R_1)}, \\
\vspace{0.2cm}
A_0=\frac{(T_1 \log R_1+T_2 \log R_2)\mu_0}{ T_1T_2\log(R_2/R_1)},\\
\vspace{0.2cm}
B_1\approx\frac{\dT}{ T_0^2\log(R_2/R_1)},\\
B_0=\frac{T_2\log R_2-T_1\log R_1}{ T_1T_2 \log(R_2/R_1)};
\eecs
\qquad\iif\;n=3, \quad
\becs
\vspace{0.2cm}
A_1\approx \frac{2R_1R_2\mu_0}{T_0(R_2-R_1) }, \\
\vspace{0.2cm}
A_0=-\frac{(R_1T_2+T_1R_2)\mu_0}{(R_2-R_1) T_1T_2},\\
\vspace{0.2cm}
B_1\approx \frac{R_2R_1\dT }{(R_2-R_1) T_0^2},\\
B_0=\frac{T_1R_2-R_1T_2}{ T_1T_2(R_2-R_1)}.
\eecs
\eeqs
Therefore, the electric flux and energy flux in radian direction are given by
\beqs \label{eq:jejuR1R2}
\becs
j_e(R_1)=-(A_1+T_0sB_1)T_0\sigma R_1^{1-n},\\
j_e(R_2)=-(A_1+T_0sB_1)T_0\sigma R_2^{1-n},
\eecs
\becs
j_u(R_1)=-[A_1\sigma s+B_1(\kappa+T_0\sigma s^2)]T_0^2 R_1^{1-n},\\
j_u(R_2)=-[A_1\sigma s+B_1(\kappa+T_0\sigma s^2)]T_0^2 R_2^{1-n},\\
\eecs
\eeqs
and hence the heat flux  in radian direction is given by
\beqs \label{eq:qR1qR2}
\becs
q(R_1)=j_u(R_1)+ \mu_0j_e(R_1),\\
q (R_2)=j_u(R_2)-\mu_0j_e(R_2).\\
\eecs
\eeqs

\subsubsection*{Power generation}
Without loss of generality we assume $\mu_0>0$. If $j_e\ge 0$, i.e.,
\beqs \label{eq:mu0ineq1}
A_1+T_0sB_1\le 0 \;\Rightarrow\;2\mu_0+ s\dT\le 0,
\eeqs
 the thermoelectric shell converts heat into electric energy.
The  electric power generated per unit volume  is given by
 \beas
\Wdot=\frac{2\sigma}{(R_2-R_1)^2}(-2\mu_0^2-\mu_0 s \dT)K_n(\frac{R_2}{R_1}),
\eeas
where $K_n(R_2/R_1)$ is a dimensionless geometric factor given by
\beas
K_n(x)=\becs
\frac{2(x-1)}{(x+1)\log x} &\iif\;n=2;\\
\frac{3x}{x^2+x+1} &\iif\;n=3.\\
\eecs
\eeas
Therefore, for fixed applied temperature difference $\dT$ the maximum power generated per unit volume is given by
\beqs \label{eq:Kn}
\Wdot_{\max}= \frac{P_f K_n(\frac{R_2}{R_1})}{4}(\frac{ \dT}{ R_2-R_1})^2\qquad \mbox{if }\quad \mu_0=-\frac{s\dT}{4},
\eeqs
where $P_f=\sigma s^2$ is the power factor of the thermoelectric material. Compared with a plate, the additional factor $K_n(\frac{R_2}{R_1})$ in the above equation is plotted in Fig.~\ref{fig:shell} (b), from which we observe that the influence is negligible for thin shells, i.e., $R_2-R_1<<R_1$.

Further, the efficiency of the conversion is given by
\beas
\eta_g=\frac{2\mu_0 j_e (R_1)}{ |j_u(R_1)|+\mu_0 j_e(R_1)}=\frac{2\sigma (-2\mu_0^2-\mu_0 s \dT)}{ |2\sigma s \mu_0T_0+(\kappa+T_0\sigma s^2) \dT |-\sigma(2\mu_0^2+\mu_0 s \dT)}.
\eeas
Upon maximizing the above efficiency again $\mu_0$ satisfying \eqref{eq:mu0ineq1}, we find the maximum efficiency is again given by \eqref{eq:etagast1} with $ZT_0=T_0s^2\sigma/\kappa$ and achieved by $\mu_0=-\dT s \sqrt{1+ZT_0}(\sqrt{1+ZT_0}-1)/2$.

\subsubsection*{Refrigeration}

Without loss of generality we assume $\dT>0$.
If $q(R_1)\ge 0$, i.e.,
\beqs \label{eq:q1ge01}
-2\sigma s \mu_0T_0-(\kappa+T_0\sigma s^2) \dT -\sigma(2\mu_0^2+\mu_0 s \dT)\ge 0,
\eeqs
the thermoelectric plate extracts heat from the low temperature side and ejects heat into the high temperature side. Since $\sigma>0$, the above inequality is possible if and only if
\beqs \label{eq:dtineq11}
(T_0+\dT/2)^2 \sigma^2s^2-2\sigma(\kappa+T_0\sigma s^2) \dT\ge 0.
\eeqs
Therefore,
\beqs \label{eq:dtineq21}
0<\dT\le \dT_{\max}:= 2T_0\frac{\sqrt{1+ZT_0}-1}{\sqrt{1+ZT_0}+1 },
\eeqs
which is the same as \eqref{eq:dtineq2}.
Further, by \eqref{eq:jejuR1R2}-\eqref{eq:qR1qR2} the cooling efficiency is given by
\beas
\eta_c =\frac{q(R_1)}{2|j_e\mu_0|},
\eeas
and upon maximizing $\eta_c$ against $\mu_0$ satisfying \eqref{eq:q1ge01} we find the maximum cooling efficiency is again given by~\eqref{eq:etacast}.

\renewcommand{\theequation}{\thesection.\arabic{equation}}
\setcounter{equation}{0}

\section{Heterogeneous thermoelectric media}

\begin{figure}[h]
\centering
\includegraphics[width=5.5in]{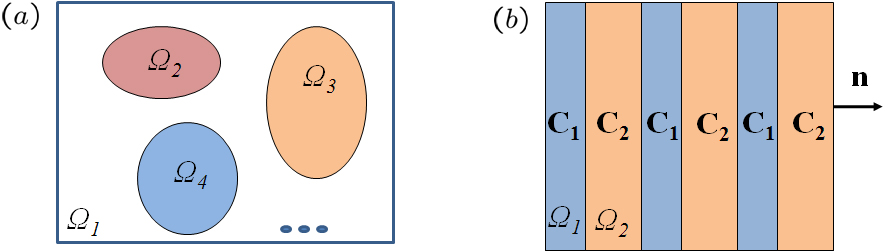}
\newline
\caption{ (a) A representive volume element of multiphase composites with the $r$th phase occupying $\Omega_r$; (b) a two-phase laminate composite of $\bfC_1$ and $\bfC_2$.}
\label{fig:2}
\end{figure}
\subsection{Effective properties of thermoelectric composites}  \label{sec:EP0}
We now consider a heterogeneous medium of $N$ individual thermoelectric phases as shown in Fig.~\ref{fig:2}(a). Let $Y=\cup_{r=1}^N \Omega_r$ be a representitive volume element or unit cell, and $\Omega_r\subset Y$  be the domain occupied by the $r$th phase $(r=1,\cdots, N)$. Suppose weak driving-forces and small variations of  $(\mu, T)$ in  the overall composite body and enforce the assumption~\eqref{eq:bfC0} for each individual phase. Note that  the electrochemical potential $\mu_0$ is  constant in the equilibrium state even if the body is heterogeneous. As for a homogeneous body,  we choose  $\mu_0=0$ and set $\bfu=(-\mu/T, 1/T)$.
From the homogenization theory (Milton 2002),  if the overall composite body is much larger than the unit cell $Y$, then the thermoelectric behavior of this composite body can be regarded as an ``effectively'' homogeneous body with an effective thermoelectric tensor $\bfC^e$. This effective tensor $\bfC^e$ can be determined by solving the unit cell problem:
\beqs \label{eq:cellproblem}
\becs
\diverg[\bfC(\bfx)(\nabla \bfv+\bfF)]=0&\oon\; Y,\\
\bfv \mbox{ is periodic}&\oon\;\partial Y,\\
\eecs
\eeqs
where $\bfv=\bfu-\bfF\bfx$,  $\bfF=\inttbar_Y \nabla \bfu\in \rz^{2\times n}$ is the average of the gradient of $\bfu$ in a unit cell,
\beqs \label{eq:bfccomp1}
&&\bfC(\bfx)= \bfC_r \qquad \iif\;\bfx\in \Omega_r,\;\; r=1,\cdots, N, \nonumber \\
&&(\bfC_r)_{1i1j}=T_0(\bfsigma_r)_{ij}, \quad (\bfC_r)_{1i2j}=(\bfC_r)_{2j1i}=T_0^2 (\bfsigma_r\bfs_r)_{ij}, \\
&&(\bfC_r)_{2i2j}= [T_0^2(\bfkappa_r+T_0\bfs_r^T\bfsigma_r \bfs_r)]_{ij}, \nonumber
\eeqs
$T_0$ is the equilibrium temperature, and $\bfsigma_r, \bfs_r, \bfkappa_r$ ($r=1,\cdots, N$) is the electric conductivity tensor, Seeback coefficient matrix, and thermal conductivity tensor of the $r$th phase, respectively.

By definition, the effective tensor $\bfC^e$, mapping  local average $\nabla \bfu$ to  local average fluxes, is such that
\beqs \label{eq:effective}
\inttbar_Y \bfJ=\inttbar_Y \bfC(\bfx) \nabla \bfu=\bfC^e \inttbar_Y \nabla \bfu=\bfC^e \bfF \qquad \forall\,\bfF\in \rz^{2\times n},
\eeqs
where $\inttbar_V\;\;$ for a domain $V$  denotes the average of the integrand over $V$.
From~\eqref{eq:effective}, we can alternately define the effective tensor $\bfC^e$ by the quadratic form
\beqs \label{eq:Ledef}
\bfF\cdot \bfC^e \bfF=\inttbar_{Y}
\bfF \cdot \bfC\left( \bfx\right) \nabla \bfu  \qquad \forall\,\bfF\in \rz^{2\times n},
\eeqs
which, in account of \eqref{eq:cellproblem} and the divergence theorem, is equivalent to
\beas
\bfF\cdot \bfC^e \bfF=\min_{\bfv\in \bbW}\bigg\{\inttbar_{Y}
 \left( \nabla \bfv+\bfF \right) \cdot \bfC\left( \bfx\right) \left( \nabla \bfv+\bfF \right) \bigg\}  \qquad \forall\,\bfF\in \rz^{2\times n}.
\eeas
Here the admissible space $\bbW$ for $\bfv$ includes the completion of all periodic continuously differentiable functions from $Y$ to $\rz^2$ in a suitable norm (i.e., $W^{1,2}$). The above variational formulation can be regarded as reminiscent of the principle of minimum entropy production of Prigozhin (1962) and will be useful for deriving bounds on the effective tensor $\bfC^e$. We will not address bounds in this paper. An excellent account of bounds on thermoelectric composites or more generally, elliptic systems with unequal dimensions of domain and range, has been presented in Bergman and Levy (1991) and Bergman and Fel (1999), where they proved that the figure of merit of two-phase composites is bounded from above by the larger one of the constituent phases if the constituent phases and the composites are isotropic.

Below we calculate the effective thermoelectric tensor of two-phase composites with microstructures of simple laminates and periodic E-inclusions. These solutions are rigorous and closed-form, giving useful guide on designing thermoelectric composites for a variety of applications.

\subsection{Effective thermoelectric properties of simple laminates} \label{sec:EPL}
As illustrated in Fig.~\ref{fig:2} (b),  we now consider a two-phase laminate composite of material property tensor $\bfC_r$,  volume fraction $\theta_r$ ($r=1,2$), and interfacial normal $\bfn$.  It is well known that the cell problem \eqref{eq:cellproblem} can be solved explicitly for laminates, see e.g. Milton (2002, page 159 and references therein). Instead of quoting the final formula, we outline below the calculations for the reader's convenience.

 Let $\bfu=\bfF\bfx+\bfv$, $\bfv$ be a solution to the cell problem~\eqref{eq:cellproblem}. Clearly the gradient field $\nabla \bfu$ and the fluxes $\bfC(\bfx)\nabla \bfu$ are constant within each laminate, and are denoted by $\bfF_r=\nabla \bfu$ on $\Omega_r$ and $\bfJ_r=\bfC_r \nabla \bfu$ on $\Omega_r$ ($r=1,2$), respectively.
 The potentials $\bfu$ being continuous across interfaces implies that $\bfF_1-\bfF_2=\bfa\otimes \bfn$ for some vector $\bfa\in \rz^2$.
 Also, we have $\theta_1 \bfF_1+\theta_2\bfF_2=\intbar_Y\nabla \bfu=\bfF$, and henceforth

\beqs \label{eq:bfF12}
\bfF_1=\bfF+\theta_2\bfa\otimes \bfn,\qquad \bfF_2=\bfF-\theta_1\bfa\otimes \bfn.
\eeqs
Further, the conservation laws across interfaces imply
\beqs \label{eq:divCF0}
(\bfJ_1-\bfJ_2)\bfn=(\bfC_1\bfF_1-\bfC_2\bfF_2)\bfn=0.
\eeqs
Inserting \eqref{eq:bfF12} into \eqref{eq:divCF0} we obtain
\beqs \label{eq:bfa1}
[(\bfC_1-\bfC_2)\bfF]\bfn+[(\theta_2\bfC_1+\theta_1\bfC_2) \bfa\otimes \bfn]\bfn=0.
\eeqs
Moreover, by definition we have
\beqs \label{eq:bfCe1}
\bfJ=\inttbar \bfC(\bfx)\nabla \bfu=\theta_1\bfC_1\bfF_1+\theta_2\bfC_2\bfF_2
=(\theta_1\bfC_1+\theta_2\bfC_2) \bfF+\theta_1\theta_2 (\bfC_1-\bfC_2)\bfa\otimes \bfn=:\bfC^e\bfF.
\eeqs
For future convenience, we introduce notations:
\beas
\bfCtld_\theta=\theta_2\bfC_1+\theta_1\bfC_2, \qquad \bfC_\theta=\theta_1\bfC_1+\theta_2\bfC_2.
\eeas
Also, let $\bfN$ be the $2\times 2$ inverse matrix of
$(\bfCtld_\theta)_{piqj} (\bfn)_i(\bfn)_i$ $(p,q=1,2)$, and $\bfS:\rz^{2\times n}\to \rz^{2\times n}$
be a symmetric tensor with its components given by
\beas
(\bfS)_{piqj}=(\bfN)_{pq}(\bfn)_{i}(\bfn)_{j}. 
\eeas
Upon solving \eqref{eq:bfa1} for the vector $\bfa$ and inserting it into \eqref{eq:bfCe1} we find
\beqs \label{eq:bfCe2}
\bfC^e=\bfC_\theta -\theta_1\theta_2 (\bfC_1-\bfC_2)\bfS (\bfC_1-\bfC_2).
\eeqs

If the constituent phases are isotropic in the sense that
\beas
\bfsigma_r=\sigma_r\bfI, \quad \bfs=s_r \bfI, \quad \bfkappa_r=\kappa_r\bfI \qquad (r=1,2),
\eeas
or equivalently (cf. \eqref{eq:bfccomp1}),
\beqs \label{eq:bfccomp2}
&&(\bfC_r)_{piqj}=(\bfA_r)_{pq}\delta_{ij},
\qquad
\bfA_r=
\bbm
T_0 \sigma_r& T_0^2\sigma_rs_r\\
T_0^2\sigma_rs_r & T_0^2\kappa_r+T_0^3 s_r^2\sigma_r\\
\ebm,
\eeqs
then equation~\eqref{eq:bfCe2} implies the effective tensor $\bfC^e$ has its components given by
\beqs  \label{eq:bfCe3}
(\bfC^e)_{piqj}=(\bfA_\theta)_{pq} \delta_{ij}-\theta_1\theta_2 (\bfn)_i(\bfn)_j (\bfA_1-\bfA_2)_{pp'} (\bfAtld_\theta^{-1})_{p'q'} (\bfA_1-\bfA_2)_{q'q},
\eeqs
where $\bfA_\theta=\theta_1\bfA_1+\theta_2\bfA_2$ and $\bfAtld_\theta=\theta_2\bfA_1+\theta_1\bfA_2$. For applied boundary conditions as shown in Fig.~\ref{fig:1}(a), we denote  relevant effective material properties by
\beas
\bfA^e:=\bbm
T_0\sigma_z&T_0^2\alpha_z\\
T_0^2\alpha_z& T_0^2\kappa'_z\\
\ebm,
\eeas
where $\sigma_z, \alpha_z, \kappa_z $ are given by \eqref{eq:sigmabfe1} with the tensor $\bfC$ replaced by $\bfC^e$ in \eqref{eq:bfCe3}. If $\bfn=\bfe_z$, i.e., the  lamination direction $\bfn$ is parallel to the applied temperature and electrochemical potential gradients,  by \eqref{eq:bfCe3} we find
\beqs \label{eq:bfAe1}
\bfA^e=(\theta_1\bfA_1^{-1}+\theta_2\bfA_2^{-1})^{-1}.
\eeqs
If $\bfe_z \cdot \bfn=0$, i.e., the  lamination direction $\bfn$ is perpendicular to the applied temperature and electrochemical potential gradients, by \eqref{eq:bfCe3} we find
\beqs \label{eq:bfAe11}
\bfA^e=\bfA_\theta= \theta_1\bfA_1+\theta_2\bfA_2.
\eeqs

We remark that, though  equations~\eqref{eq:bfAe1} and \eqref{eq:bfAe11} resemble the classic mixture rules for simple laminates,    equation~\eqref{eq:bfAe1} (resp. \eqref{eq:bfAe11}) implies that the effective electric (resp. thermal)  conductivity of thermoelectric composites  is in general not given by the familiar  mixture rule: $(\theta_1/\sigma_1+\theta_2/\sigma_2)^{-1}$ (resp. $\theta_1\kappa_1+\theta_2\kappa_2$), for the coupling between electric and thermal fluxes or the nonzero off-diagonal components in $\bfA_1$ and $\bfA_2$.

\subsection{Closed-form solutions for periodic E-inclusions} \label{sec:EPE}
Particulate heterogenous media are frequently encountered for which the estimates \eqref{eq:bfCe2} and \eqref{eq:bfCe3} are not suitable for their obvious dissimilarity in microstructure. It is therefore desirable to rigorously solve the unit cell problem~\eqref{eq:cellproblem} for microstructures resembling a particulate composite and obtain its effective material properties. This is usually achieved by using the Eshelby's solution for ellipsoidal inclusions in the dilute limit where the interactions between various particles are neglected (Eshelby 1957; Mura 1987). For finite volume fractions and to account for mutual interactions between particles, we consider recently found geometric shapes called periodic E-inclusions, for which the unit cell problem~\eqref{eq:cellproblem} can be solved in a closed-form as for ellipsoidal inclusions in the dilute limit.

We consider two-phase thermoelectric composites with their thermoelectric tensors given by
 \beas
 \bfC(\bfx)=\bfC_r\qquad \iif\;\bfx\in \Omega_r, r=1,2,
 \eeas
 where $\Omega_2$ is referred to as the inclusion (occupied by a second phase thermoelectric particle), $\Omega_1$ represents the matrix (continuous) thermoelectric phase, and $\bfC_r$ ($r=1,2$) are of form \eqref{eq:bfccomp2}. We remark that many but not all anisotropic thermoelectric tensors can be transformed into this form under linear transformations.

  Below we solve explicitly the unit cell problem \eqref{eq:cellproblem} for $\Omega_2$ being a periodic E-inclusion following the well-known Eshelby equivalent inclusion method. First, we recall that a periodic E-inclusions, by definition, is such that the overdetermined problem  (Liu {\em et al.} 2007; Liu {\em et al.} preprint)
\begin{equation}
\left \{
\begin{array}
[c]{ll}
\nabla^2 \phi=\theta-\chi_{\Omega_2}&\oon\;Y,\\
\nabla \nabla \phi=-(1-\theta)\Qbf  &\oon\;{\Omega_2},\\
\mbox{periodic boundary conditions} &\oon\;\partial Y,\\
\end{array}
\right.
\label{eq:potential}
\end{equation}
  admits a solution, where $\theta =|{\Omega_2}|/|Y|$ is the volume fraction of the inclusion, and the shape matrix  $\Qbf$ is necessarily symmetric positive semi-definite with $\Tr(\Qbf)=1$.  Then, by Fourier analysis or Green's functions, we find solutions to the following {\em homogeneous} counterpart of \eqref{eq:cellproblem}
\begin{equation}
\left \{
\begin{array}
[c]{ll}
\nabla \cdot \left[\bfC_1 \nabla \bfv +\mathbf{\Sigma}^\ast\chi_{\Omega_2} \right]=0&\oon\;Y,\\
\mbox{periodic boundary conditions} &\oon\;\partial Y
\end{array}
\right.
\label{eq:homo}
\end{equation}
satisfy that, for any $\mathbf{\Sigma}^\ast\in \rz^{2\times 3}$, the gradient field $\nabla \bfv$ is uniform on $\Omega_2$
and is given by
\begin{equation}
\begin{array}
[l]{ll}
\nabla \bfv=-(1-\theta )(\bfA_1)^{-1}\mathbf{\Sigma}^\ast \bfQ=: -(1-\theta )\bfR\mathbf{\Sigma}^\ast &\qquad \oon\;\Omega_2,
\end{array}
\label{eq:bfvxi}
\end{equation}
where the components of the tensor $\bfR:\rz^{2\times n}\to \rz^{2\times n}$ are given by
\beqs
(\bfR)_{piqj}=\left( \bfA_1 \right)^{-1}_{pq} (\bfQ)_{ij}\qquad (p,q=1, 2; i,j=1,\cdots, n).
\label{eq:bfR1}
\eeqs
We now consider the inhomogeneous problem (\ref{eq:cellproblem}). Based on  the
equivalent inclusion method we observe that the solution to \eqref{eq:cellproblem} is identical to that of \eqref{eq:homo} if the average applied  field $\Fbf$ for \eqref{eq:cellproblem} and the ``eigenstress" $\mathbf{\Sigma}^\ast$ for \eqref{eq:homo} are related by (Liu {\em et al.} preprint)
\beqs
\dbfC\Fbf=(1-\theta ) \dbfC\Rbf\mathbf{\Sigma}^\ast-\mathbf{\Sigma}^\ast=[(1-\theta )\dbfC\Rbf-\IIbf]\mathbf{\Sigma}^\ast,
\label{FPeq}
\eeqs
where $\dbfC=\bfC_1-\bfC_2$ and $\IIbf:\rz^{2\times n}\to \rz^{2\times n}$ is the identity mapping.

To calculate the effective tensor of the composite, by \eqref{eq:Ledef} and \eqref{eq:bfvxi}
 we find that the effective tensor $\bfC^e$ satisfies
\beas
&&\Fbf\cdot \bfC^e \Fbf=\inttbar_{Y}
\Fbf \cdot (\bfC_1-\dbfC \chi_{\Omega_2})\left( \nabla \ubf+\Fbf \right) d\xbf\\
&&\hspace{1.5cm}=\Fbf\cdot \bfC_1\Fbf-\theta \Fbf\cdot \dbfC[-(1-\theta )\Rbf\mathbf{\Sigma}^\ast+ \Fbf].
\eeas
By \eqref{FPeq} we rewrite the above equation as
\beas
\Fbf\cdot \bfC^e \Fbf
=\Fbf\cdot \bfC_1\Fbf+ \theta \Fbf\cdot \mathbf{\Sigma}^\ast.
\eeas
Further, it can be shown that the tensor $(1-\theta )\dbfC\Rbf-\IIbf$ is invertible for generic cases and the above
equation implies
\beqs
\bfC^e=\bfC_1+\theta  [(1-\theta )\dbfC\Rbf-\IIbf]^{-1}\dbfC,
\label{eq:Leffective}
\eeqs
which is our closed-form formula of effective tensors for two-phase thermoelectric composites. 

We remark that  formula~\eqref{eq:Leffective} is a rigorous  prediction to the effective properties of thermoelectric composites with microstructures of periodic $E$-inclusions and there is no phenomenological parameters in~\eqref{eq:Leffective}.  Further, the shape matrix $\bfQ$ can be any  positive semi-definite matrix with $\Tr(\Qbf)=1$ and the volume fraction $\theta $ can be any number in $(0,1)$, which allows for modeling composites with a variety of textures and volume fractions of inhomogeneities. In particular, if the shape matrix $\Qbf$ is a rank-one matrix $\bfn\otimes \bfn$ for a unit vector $\bfn$, the inclusion degenerates to a laminate perpendicular to $\bfn$ and \eqref{eq:Leffective} recovers the formula~\eqref{eq:bfCe3} for simple laminated composites whereas if $\bfQ=\bfI/n$ we have
 \beqs \label{eq:bfAe2}
( \bfC^e)_{piqj}=(\bfA^e)_{pq}\delta_{ij},\qquad \bfA^e=\bfA_1+\theta [(1-\theta ) \dbfA \bfA_1^{-1}/n-\bfI]^{-1}\dbfA.
 \eeqs

\subsection{Applications to the design of thermoelectric composites} \label{sec:EPA}

\begin{figure}[t]
\centering
\includegraphics[width=6in]{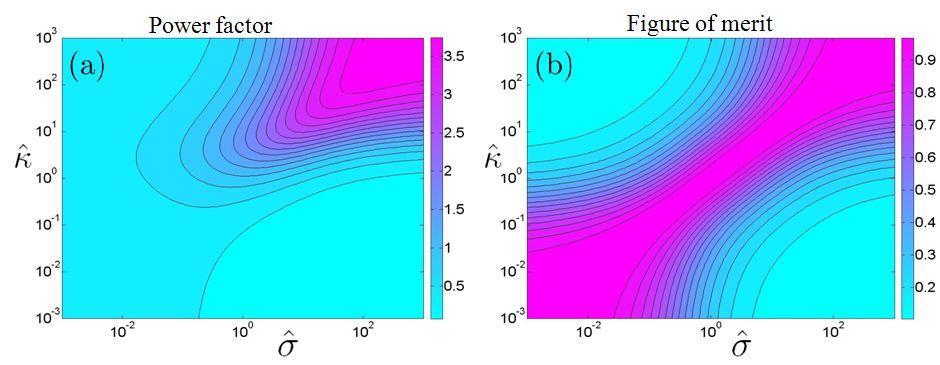}
\caption{ (a) Contours of normalized effective power factors $P^e_f/P^b_f$, and (b) contours of normalized effective power factors $Z^eT_0$. The horizontal and vertical axes represent $(\sigmahat, \kappahat)$ (cf., \eqref{eq:sskhat}), and $\shat$ is such that the figure of merit of the $a$-phase is the same as the $B$-phase. The composites consist of $a$-phase periodic E-inclusions  in $B$-phase matrix with isotropic shape matrix and volume fraction $\theta_a=0.5$.}
\label{fig:ff4}
\end{figure}

The closed-form formula~\eqref{eq:Leffective} are useful for engineering the efficiency and power factor of thermoelectric composites. To this end, we choose p-type Bismuth-telluride (Bi$_{0.25}$ Ti$_{0.75}$ Te$_3$ doped with 6 wt\% excess Te) as the base material (referred to as the $B$-phase). The thermoelectric properties of this p-type Bismuth-telluride
are listed below (Yamashita and Odahara 2007):
\beas
 \sigma_b=0.326\times 10^5 (\Omega m)^{-1}, \qquad  s_b=245.0 \mu V/K, \qquad  \kappa_b= 0.559 W/m K,
 \eeas
 by which the power factor and figure of merit can be easily calculated and are given by
 \beas
  \; P_f^b=2.0\times 10^{-3} W/mK^2,\qquad Z_bT_0=1.04.
 \eeas
 Suppose the base material be mixed with a second isotropic thermoelectric material (referred to as the $A$-phase) with
\beqs \label{eq:sskhat}
 \sigma_a=\sigmahat \sigma_b, \quad s_a= \shat s_b, \quad \kappa_a=\kappahat \kappa_b,
 \eeqs
 where $\sigmahat,\; \shat,\;\kappahat$ are, respectively, dimensionless scaling factors of electric conductivity, Seebeck coefficient, and thermal conductivity. Below we assume a family of shape matrices $\bfQ$ parameterized by a scalar $\omega\ge 1$:
\beqs \label{eq:bfQomega}
\bfQ(\omega)=\bbm
\frac{1}{2+\omega}&0&0\\
0& \frac{1}{2+\omega}&0\\
 0&0&\frac{\omega}{2+\omega}\\
\ebm.
\eeqs
We remark that if $\omega\to0$ and $\theta>0$,  the periodic E-inclusion approaches to a fiber in $\bfe_z$-direction,
and if $\omega\to \infty$ and $\theta>0$, the periodic E-inclusion approaches to a laminate in $\bfe_z$-direction.
From \eqref{eq:Leffective} we can write the effective power factor and figure of merit along $\bfe_z$-direction as functions of
$\omega, \theta, \sigmahat, \shat, \kappahat$:
\beas
&&P_f^e=P_f^e(\omega, \theta, \sigmahat, \shat, \kappahat),\\
&&Z^eT_0=Z^eT_0(\omega, \theta,  \sigmahat, \shat, \kappahat).
\eeas
Below we investigate numerically how  the effective power factor and figure of merit depend on these parameters.

Since it is hard to improve the figure of merit of thermoelectric materials, we first vary $\sigmahat, \kappahat$ between $(10^{-3}, 10^3)$ and choose  $\shat=(\sigmahat/\kappahat)^{1/2}$  such that the figure of merit of $A$-phase is the same as that of $B$-phase. We assume the microstructure of composites are $A$-phase particles of the shape of periodic E-inclusion embedded in $B$-phase matrix. The volume fraction of inclusions is fixed at $0.5$ and the shape matrix is given by \eqref{eq:bfQomega} with $\omega=1$. Figure~\ref{fig:ff4} (a) shows contours of normalized effective power factors $P^e_f/P^b_f$ whereas Fig.~\ref{fig:ff4} (b) shows contours of effective  figure of merit $Z^eT_0$. From these figures we observe that for fixed thermal conductivity there exists an optimal electric conductivity for maximizing the effective power factor and vice versa. However, the power factor cannot be significantly improved (within five fold). Also, as shown in Bergman and Levy (1991) the effective figure of merit cannot exceed that of the constituent phases. Moreover, if electric and thermal conductivity are uniformly scaled, i.e., $\sigmahat=\kappahat$, the effective figure of merit is the same as that of the constituent phases. This remains true even if $\shat=\kappahat=0$, which implies that porous media of high figure of merit materials would likely keep the property of high figure of merit  (Inoue {\em et al} 2008).  Additionally, we remark that similar qualitative results are observed when  volume fraction $\theta$ and shape factor $\omega$ are varied and when inclusion material and matrix material are interchanged.

\begin{figure}[t]
\centering
\includegraphics[width=6in]{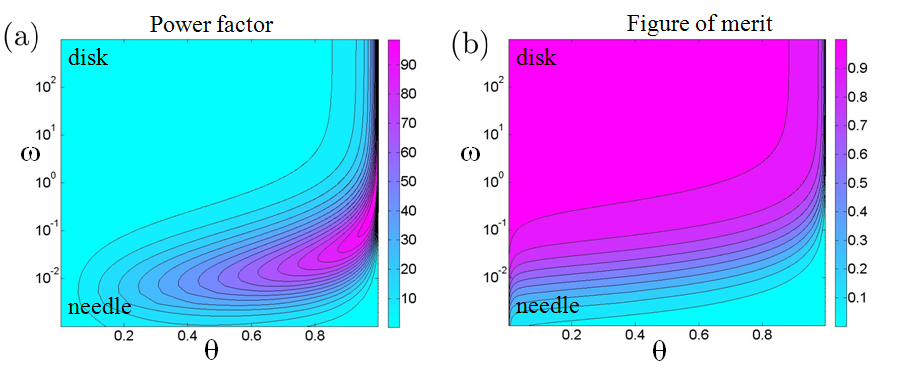}
\caption{ (a) Contours of normalized effective power factors $P^e_f/P^b_f$, and (b) contours of normalized effective power factors $Z^eT_0$. The horizontal and vertical axes represent volume fraction and shape factor $(\theta_a, \omega)$ (cf., \eqref{eq:bfQomega}), and the composites consist of $A$-phase periodic E-inclusions  in $B$-phase matrix. The material properties of $A$-phase takes values of copper. }
\label{fig:ff5}
\end{figure}

Next we consider composites of copper ($A$-phase) periodic E-inclusions embedded in $B$-phase matrix. The scaling factors are given by $\sigmahat=2.0\times 10^3$, $\shat=7.8\times 10^{-3}$ and  $\kappahat=7.17\times 10^2$  for copper (Yamashita and Odahara 2007). Figure~\ref{fig:ff5} (a) \& (b) show respectively contours of effective power factor $P^e_f/P^b_f$  and effective figure of merit $Z^eT_0$, where the horizontal and vertical axes represent volume fraction and shape factor $(\theta, \omega)$. Qualitatively, the shape of periodic E-inclusion is like a needle (disk) when $\omega<<1$ ($\omega >>1$). From Fig.~\ref{fig:ff5} (a) we observe that the effective power factor increases as the volume fraction of $A$-phase increases. In contrast, there exists an optimal value of $\omega$ for maximizing power factor. As volume fraction $\theta$ increases from $0.1$ to $0.9$, the optimal shape gradually changes from needle-like shapes to disk-like shapes.  Also, the power factor can be significantly improved (up to $100$ fold). From Fig.~\ref{fig:ff5} (b) we observe that disk-like shapes in general have higher figure of merit than needle-like shapes. This dependence is however weak in the sense that the change of effective figure of merit is within $10 \sim 20\%$ even if $\omega$ varies between $(0.1, \;10^3)$. Therefore, the trade-off between power factor and figure of merit often makes shapes with  significantly improved power factor but a sightly lower figure of merit more desirable.  Figure~\ref{fig:ff5} (a) \& (b) provide valuable information for choosing optimal volume fractions and shapes of copper particles to fulfill  practical requirements. Finally we remark that interchanging inclusion and matrix phase materials yields undesirable lower power factor and figure of merit.

\section{Summary and discussion} \label{sec:SD}

We have developed  systematically a continuum theory for thermoelectric bodies following the general framework of continuum mechanics and conforming to basic thermodynamic laws. Under reasonable assumptions of small variations of electrochemical potential and temperature, the resulting boundary value problem is a linear system of partial differential equations which can be conveniently solved to determine  all relevant local fiends. We have also applied the theory to predict the effective properties of thermoelectric composites. In particular, explicit formula of the effective thermoelectric tensors have been calculated for simple microstructures of laminates and periodic E-inclusions. These explicit formula facilitate investigations of how practically important material properties such as power factor and figure of merit depend on the volume fractions, shapes of inclusions and material properties of constituent phases.

\newcommand{\bfadot}{{\dot{\bfa}}}
\renewcommand{\theequation}{A.\arabic{equation}}
\setcounter{equation}{0}

\section*{Appendix A: Onsager's reciprocal relations and their applications to thermoelectric materials}

The original Onsager's derivation of the reciprocal relations of thermoelectric materials  is somewhat elusive to researchers in the community of continuum mechanics. Following Casimier (1945)  we outline below the premises for deriving the reciprocal relations for thermoelectric materials whereas detailed derivations can be found in Onsager (1931) and Casimier (1945).
\subsection*{General systems in irreversible processes}
We consider a closed isolated system. Assume the system is described by a set of macroscopic thermodynamic  variables $a^i$ ($i=1,\cdots, N$) whose values, without loss of generality, are assumed to be zero in the equilibrium state. Although the system is in the equilibrium state, the system can fluctuate, which is characterized by time-dependent random variables $\bfa(t)=(a^1,\cdots, a^N)^T$.

The derivation of Onsager's reciprocal relations requires the following premises:
\begin{enumerate}
  \item The entropy in a state  $\bfa$ deviating from the equilibrium state, to the leading order, is given by
\beqs \label{eq:Sexp}
S(\bfa)=S_0-\half \bfa\cdot \bfS \bfa,
\eeqs
where the constant $N\times N$ symmetric matrix $ \bfS$ is  a property of the system, and positive definite since entropy is necessarily an maximum in the equilibrium state. From the definition of entropy (Kittel and Kroemer 1980, page 40), the number of microscopic admissible states of the system is given by
$
g(\bfa)\propto \exp(-\frac{1}{2k_B} \bfa\cdot \bfS \bfa),
$
and the probability density function is given by
\beas
W(\bfa)=\frac{1}{C}\exp(-\frac{1}{2k_B} \bfa\cdot \bfS \bfa),\eeas
where 
 $C=\sqrt{(2\pi)^Nk_B^N/\det \bfS} $ is a normalization constant such that $\int_{\rz^N} W(\bfa) d\bfa=1$.
 Let $\bfb=\bfS \bfa$. Direct integration shows that
\beqs \label{eq:Eba}
E[\bfb\otimes \bfa]=\int_{\rz^N} \bfb\otimes \bfa W(\bfa)d\bfa=k_B \bfI,
\eeqs
where $\bfI$ is the $N\times N$ identity matrix.
\item Also, we assume the {\em ergodicity} of all admissible microscopic states such  that the expectation of a quantity $f(\bfa)$ is exactly equivalent to a time averaging:
\beqs \label{eq:Atave}
E[f(\bfa(t))]=\lim_{T\to +\infty} \frac{1}{T}\int_0^T f(\bfa(t))dt.
\eeqs

\item Let
\beas
\bfatld(t,\tau)=E[\bfa(t+\tau)|\bfa(t)]
\eeas
be the conditional expectation characterizing the probability of finding the system in states $\bfa(t+\tau)$ at time $t+\tau$ while knowing the system is in state $\bfa(t)$ at time $t$.
We consider only state variables $\bfa$ which are even functions of velocity (Casimir 1945).
Then the microscopic reversibility implies
\beqs \label{eq:MicroR}
\bfatld(t, \tau)=\bfatld(t, -\tau).
\eeqs

Therefore,
\beqs \label{eq:Eaa0}
 E[\bfa(t) \otimes \bfa(t+\tau)]=
E[\bfa(t) \otimes \bfatld(t, \tau)]=E[\bfa(t) \otimes \bfatld(t, -\tau)]=E[\bfa(t+\tau) \otimes \bfa(t)],
\eeqs
where the first equality follows from  the general probability theory (see e.g.  Evans 2002,  page 31), the second equality follows from \eqref{eq:MicroR}, and the last equality follows from \eqref{eq:Atave}.

\item If for irreversible nonequilibrium processes the evolution of the system follows  the macroscopical kinetic law for a constant $N\times N$ matrix~$\bfM$:
\beqs \label{eq:aSb}
\bfadot(t)=\bfM \bfb(t),
\eeqs
then the fluctuation (in the equilibrium state) follows the above law as well in the sense that
\beqs \label{eq:Eaa}
E[\bfa(t+\tau)- \bfa(t)|\bfa(t)]=\tau \bfM\bfb(t).
\eeqs
The physical meaning and conditions for the above hypothesis have been elaborated in Casimir (1945).
\end{enumerate}
Therefore, by \eqref{eq:Eba} we have
\beas
&&E[\bfa\otimes (\bfa(t+\tau)-\bfa(t))]=
E[\bfa\otimes E[\bfa(t+\tau)-\bfa(t)| \bfa(t)]]=\tau E[ \bfa\otimes \bfM \bfb(t)]=\tau k_B \bfM^T,\\
&&E[ (\bfa(t+\tau)-\bfa(t))\otimes \bfa]=
E[E[\bfa(t+\tau)-\bfa(t)| \bfa(t)]\otimes \bfa ]=\tau E[ \bfM \bfb(t)\otimes \bfa(t)]=\tau k_B \bfM,
\eeas
and henceforth, by \eqref{eq:Eaa0}, arrive at
\beqs \label{eq:reciprocal}
\bfM^T=\bfM.
\eeqs

\subsection*{Applications to thermoelectric materials}

To apply the above general calculations to thermoelectric materials, we choose $\bff_1=(-\nabla \mu/T, -\nabla T/T)^T$ as the thermodynamic variables: $\bfa=\bff_1$, and assume that the system has unit volume. Taking the time derivative of \eqref{eq:Sexp} we obtain
\beas
\gamma=\frac{d S(\bfa)}{dt}=- \bfa \cdot \bfS \bfadot,
\eeas
where $\gamma$ is the entropy production rate per unit volume. From \eqref{eq:gammajf} we have
\beas
\bfj_1=(\bfj_e, \bfj_s)=- \bfS \bfadot, \qquad
\eeas
Therefore, if $\bfadot$ and $\bfb$ are related by \eqref{eq:aSb}, then
\beas
\bfj_1=- \bfS \bfadot=-\bfS \bfM\bfb=-\bfS\bfM\bfS \bfa=-\bfS\bfM\bfS \bff_1.
\eeas
Comparing the above equation with \eqref{eq:constituitive} we have $\bfL_1=-\bfS\bfM\bfS$, and henceforth
\beas
\bfL_1^T=\bfL_1
\eeas
since $\bfM^T=\bfM$ and $\bfS^T=\bfS$.

{\bf Acknowledgements:} The author gratefully acknowledges the support of NSF under grant no. CMMI-1101030 and AFOSR (YIP-12). This work was mainly carried out while the author held a position at University of Houston; the support of UH and TcSUH is gratefully acknowledged.

\end{document}